\documentclass[a4paper,12pt]{article}
\usepackage[usenames]{color}
\usepackage{array}
\usepackage{cite}
\usepackage{amsmath}
\usepackage{amssymb}
\usepackage{amsthm} 
\usepackage{mathrsfs} 
\usepackage[dvipdfmx]{graphicx}
\usepackage[framemethod=tikz]{mdframed}
\usepackage{latexsym} %

\newcommand{\rack}{\triangleright}
\newcommand{\prack}{\trianglerighteq}
\newcommand{\cat}{\rightrightarrows}

\theoremstyle{definition}
\newtheorem*{define}{Definition}
\newtheorem*{proposition}{Proposition}
\newtheorem*{lemma}{Lemma}
\newtheorem*{theorem}{Theorem}
\newcommand{\calL}{{\cal L}}
\newcommand{\cbral}{[\![}
\newcommand{\cbrar}{]\!]}
\newcommand{\vbracket}[2]{\cbral #1,\,#2\cbrar}

\newcommand{\bracket}[2]{( #1,\,#2)}
\def\bR{{\mathbb{R}}}

\newcommand{\Map}{{\rm Map}}

\newcommand\wM{\mathcal{M}}

\def\tilx{\widetilde{X}{}}
\def\tily{\widetilde{Y}{}}

\def\tila{\widetilde{\alpha}{}}
\def\tilb{\widetilde{\beta}{}}

\begin{document}
\begin{titlepage}
\null
\begin{flushright}
June, 2020
\end{flushright}
\vskip 3cm
\begin{center}
{\Large \bf 
Global Aspects of Doubled Geometry
 \\
\vspace{0.5cm}
and Pre-rackoid 
}
\vskip 1.5cm
\normalsize
\renewcommand\thefootnote{\alph{footnote}}

{\large
Noriaki Ikeda\footnote{nikeda(at)se.ritsumei.ac.jp}
and 
Shin Sasaki\footnote{shin-s(at)kitasato-u.ac.jp}
}
\vskip 0.7cm
  {\it
   ${}^{\rm{a}}$Department of Mathematical Sciences, Ritsumeikan University \\
   Kusatsu, Shiga 525-8577, Japan, \\
   \vskip 0.3cm
   ${}^{\rm{b}}$Department of Physics,  Kitasato University \\
   Sagamihara 252-0373, Japan
}
\vskip 1.7cm
\begin{abstract}
The integration problem of a C-bracket and a Vaisman (metric, pre-DFT)
 algebroid which are geometric structures of double field theory (DFT) is
 analyzed. 
We introduce a notion of a pre-rackoid as a global group-like object for an infinitesimal algebroid structure.
We propose that several realizations of pre-rackoid structures.
One realization is that elements of a pre-rackoid are defined by cotangent paths along doubled foliations in a para-Hermitian manifold.
Another realization is proposed as a formal exponential map of the algebroid of DFT.
We show that the pre-rackoid reduces to a rackoid that is the integration of the Courant algebroid when the strong constraint of DFT is imposed.
Finally, for a physical application, we exhibit an implementation of
 the (pre-)rackoid in a three-dimensional topological sigma model.

\end{abstract}
\end{center}

\end{titlepage}

\newpage
\setcounter{footnote}{0}
\renewcommand\thefootnote{\arabic{footnote}}
\pagenumbering{arabic}
%
\section{Introduction} \label{sct:introduction}
Double field theory (DFT) \cite{Hull:2009mi},
based on the doubled formalism developed in \cite{Siegel:1993th, Siegel:1993bj},
 is a gravity theory that inherits T-duality in
string theory. 
DFT is defined in a $2D$-dimensional doubled spacetime where
the Kaluza-Klein and the string winding modes are realized by the doubled
coordinates $x^M = (x^{\mu},\tilde{x}_{\mu})$. In this formalism, T-duality is
implemented as a global $O(D,D)$ symmetry in the doubled space.
In DFT, the physical $D$-dimensional spacetime is defined through the imposition
of the strong constraint. Under the strong constraint, the DFT action is
reduced to one of supergravity of NSNS sectors in an appropriate frame.
Such kind of structure is naturally incorporated in a para-Hermitian
manifold $\mathcal{M}$ \cite{Vaisman:2012ke, Vaisman:2012px, Freidel:2017yuv,
Freidel:2018tkj}. 
It has been shown that the $D$-dimensional physical spacetime appears as a leaf in doubled foliations of $\mathcal{M}$. 

The gauge symmetry of DFT originates from the diffeomorphism and the
$U(1)$ gauge symmetry of the NSNS $B$-field.
The transformations of DFT fields are generated by vector fields in the doubled
space and they are governed by a C-bracket.
An algebra based on the C-bracket is known to be a Vaisman (metric or
pre-DFT) algebroid \cite{Vaisman:2012ke, Svoboda:2018rci, Chatzistavrakidis:2018ztm}.
This is a generalization of a Courant algebroid \cite{Courant:1990} that
plays an important role in generalized geometry.
Indeed, when the strong constraint is imposed on any vector field and
function in $\mathcal{M}$, the Vaisman algebroid reduces to the Courant
algebroid \cite{Hull:2009zb}.
Both algebroids exhibit local structures of the tangent bundle
$T\mathcal{M}$ of the para-Hermitian manifold.

An important aspect of these algebroids involves their doubled structures.
For example, it has been shown that the Courant algebroid is composed
by a Drinfel'd double of Lie bialgebroids \cite{Liu:1997}.
An analogous result has been obtained even for a Vaisman algebroid.
In this picture, the geometric origin of the strong constraint in DFT is traced back to a consistency condition of the Drinfel'd double for Lie bialgebroids \cite{Mori:2019slw}.
The doubled aspects appearing in the contexts of mathematical and physical
sides of DFT shed light on the deep understanding of nature of T-duality.
Among other things, the structure of the Drinfel'd double of Lie bialgebras is a key ingredient of the Poisson-Lie T-duality
\cite{Klimcik:1995ux, Klimcik:1995jn, Klimcik:1995dy}.
This is a generalization of ordinary T-duality with Abelian isometries
and its physical applications have been studied \cite{Klimcik:2002zj,
Klimcik:2008eq, Hassler:2017yza, Marotta:2018myj, Sakatani:2019jgu}.
The Poisson-Lie T-duality is interpreted as a freedom of choices for Manin
triples in the Drinfel'd double of the Lie bialgebras.
If there are several Manin triples, one can generalize the duality to
plurality \cite{VonUnge:2002xjf}. 
Recent developments along this line include \cite{Sakatani:2019zrs, Malek:2019xrf, Sakatani:2020iad, Hlavaty:2020pfj}.

It is well known that the Lie bialgebra is the infinitesimal object of
the Poisson-Lie group. Conversely, a Lie bialgebra is
integrated to a Poisson-Lie group. Obviously, the Poisson-Lie T-duality
is named after this group structure.
One can imagine that this picture may be generalized to algebroids even in the absence of group structures.
Indeed, a Lie algebroid is an infinitesimal object of a Lie groupoid.
By the same way, a Lie bialgebroid is an infinitesimal object of a
Poisson groupoid \cite{Mackenzie:1994}. 
It is known that the Drinfel'd double of Lie bialgebroids gives a Courant algebroid.
Manin triples are defined by a Dirac structure of a Courant algebroid
\cite{Liu:1997}.
It is therefore conceivable that there is a groupoid-like structure
defined by the integration of the Courant algebroid.
However, it is a highly non-trivial task to determine a global,
group-like object for a given algebroid.
This is because there is no analogue of the Lie's third theorem
associated with a Lie algebroid and its global counterpart.
Finding a group-like object from a given algebra is known as the {\it
coquecigrue problem} \cite{Loday:1993}.
The coquecigrue problem for the Courant algebroids has been studied
intensively in various contexts \cite{Severa:2001, Sheng:2011, Bland:2012, Mehta:2013}.

The purpose of this paper is to investigate a global structure of the
doubled spacetime and examine the geometric meaning of the strong
constraint in DFT.
To this end, we work on the coquecigrue problem of the Vaisman
algebroid which is a natural local structure appearing in a para-Hermitian manifold. 
We remind ourselves that the standard Courant
algebroid is recognized as a Leibniz algebroid \cite{Liu:1997}.
It is known that an integration of a Leibniz algebra is given by a {\it rack}
\cite{Kinyon:2004, Covez:2010, Bordemann:2016}.
Racks and associated quandles were first proposed in the contexts of
knot theory \cite{Carter:2010}. 
They are sets equipped with binary operations satisfying certain axioms.
Notably, racks are generalizations of groups. 
Roughly speaking, racks are group-like objects that are based on the conjugation instead of the product in the sense of ordinary groups.
Accordingly, it seems plausible that an integration of a Leibniz
algebroid is given by a groupoid-like counterpart of the rack -- the
{\it rackoid}.
Indeed, the authors in \cite{Laurent-Gengoux:2018zoz} proposed that an
integration of the standard Courant algebroid is given by a rackoid.
This rackoid structure is defined by cotangent paths associated with
underlying paths in the base space.
The standard Courant algebroid is shown up as an infinitesimal (or
tangent bundle) structure of the cotangent path rackoid.
We will generalize this picture to the Vaisman algebroid and
look for a groupoid-like structure, which we call the {\it pre-rackoid}, in
the doubled spacetime in DFT.

We also propose another method, a formal exponential map of a Vaisman
algebroid.We consider an exponential of the adjoint operation. It is
called a formal (pre-)rackoid. This idea is directly related to one of a
large gauge transformations of DFT \cite{Hohm:2012gk}.

As an explicit realization and applications, we discuss the
(pre-)rackoid structures in a topological sigma model.
A three dimensional topological sigma model with the structure of a Courant algebroid called the Courant sigma model is known \cite{Ikeda:2002wh, Roytenberg:2006qz}. This model is useful to analyze the integration of the Courant algebroid to Lie rackoid. Wilson lines in the model realize a Lie rackoid structure.
The doubled geometry version of topological sigma model has been proposed 
\cite{Kokenyesi:2018xgj}. 
See also \cite{Chatzistavrakidis:2018ztm, Chatzistavrakidis:2019rpp}.
We will show that the topological sigma model of doubled geometry gives a realization of a (formal) pre-rackoid.

The organization of this paper is as follows.
In the next section, we give the definition of the Vaisman algebroid and
its relations with the Courant algebroid.
In section \ref{sct:rackoids}, we introduce the notion of the rack and
the rackoid. 
Mathematical definitions of these structures are presented.
In section \ref{sct:pre-rackoids}, we first focus on the cotangent path rackoid 
discussed in \cite{Laurent-Gengoux:2018zoz}.
This gives an integration of the standard Courant algebroid.
We then generalize this to the doubled cotangent path defined in the doubled foliations of the
para-Hermitian manifold $\mathcal{M}$.
We show that the integration of the Vaisman
algebroid based on the C-bracket is given by a generalization of
rackoid, namely, the pre-rackoid.
We show that the rack-like product based on the doubled cotangent path defines the pre-rackoid.
In this picture, the strong constraint in DFT is an sufficient
condition of the self-distributivity of the rack product.
We will see that this is re-organized as the quantum Yang-Baxter equation for the rack
action.
In section \ref{sct:formalrackoid}, we show another way to provide the
(pre-)rackoids. We introduce formal exponential maps of the adjoint
action to define the (pre-)rack product.
This procedure gives a formal integration of the Courant and
the Vaisman algebroids.
In section \ref{sct:sigma_models}, we discuss a sigma model
implementation of the (pre-)rackoid structures.
We show that the (pre-)rackoids associated with the
Vaisman and the Courant algebroids are realized as Wilson loops in the sigma models.
Section \ref{sct:conclusion} is the conclusion and discussions.

\section{Leibniz, Courant and Vaisman algebroids}
\label{sct:algebroids}
In this section, we introduce the Courant and the Vaisman algebroids.
The latter appears in para-Hermitian manifolds which are natural arenas
of the doubled spacetime in DFT.
Before the definition of the Courant and the Vaisman algebroids, let us first remind the definition of a Leibniz algebra.
\begin{define}[(Left) Leibniz algebra (Loday algebra)]
\label{def:Leibniz_algebra}
A (left) \textit{Leibniz algebra (Loday algebra)} $\mathfrak{g}$ is defined as a module over a ring $R$ equipped with a bilinear map $[\cdot,\cdot]$ (the Leibniz bracket) on $\mathfrak{g}$ satisfying the following left
 Leibniz identity:
\begin{align}
[a,[b,c]] = [[a,b],c] + [b,[a,c]], \quad \text{for} \
 {}^{\forall} a,b,c \in \mathfrak{g}.
\label{eq:Leibniz_identity}
\end{align}
A right Leibniz algebra is defined similarly.
\end{define}
Note that the Leibniz bracket $[\cdot,\cdot]$ is 
not necessarily skew-symmetric.
When the Leibniz bracket $[\cdot,\cdot]$ is skew-symmetric, it becomes a Lie algebra.
The notion of the Leibniz algebra is easily generalized to the ones of algebroids.
\begin{define}[(Left) Leibniz algebroid]
\label{def:Leibniz_algebroid}
A (left) \textit{Leibniz algebroid} is a triple $(E,[\cdot,\cdot]_D,\rho)$, where $E \xrightarrow{\pi} M$ is a vector bundle over a smooth manifold $M$, 
 $[\cdot,\cdot]_D: \Gamma (E) \times \Gamma (E) \to \Gamma (E)$ is
a Leibniz bracket satisfying the Leibniz identity \eqref{eq:Leibniz_identity},
$\rho: E \to TM$ is a bundle map called the anchor map, and
 $[\cdot,\cdot]_D$ and $\rho$ satisfy the following 
relations:
\begin{align}
&
\rho ([e_1,e_2]_D) = [\rho (e_1), \rho (e_2)]_{TM}, 
\notag \\
&
[e_1, f e_2]_D = f [e_1, e_2]_D + (\rho (e_1) \cdot f) e_2,
\notag \\
& 
\text{for any } e_i \in \Gamma (E) \text{ and } f \in C^{\infty}(M).
\end{align}
Here $[\cdot,\cdot]_{TM}$ is the Lie bracket of vector fields on $TM$.
We note that the first relation can be omitted since it is obtained through the second one and
 the Leibniz identity \eqref{eq:Leibniz_identity}. 
In the Leibniz algebroid, the bracket is called the Dorfman bracket.
\end{define}
The Leibniz algebroid is a generalization of the Lie algebroid and the
Leibniz algebra.
When the bracket $[\cdot,\cdot]_D$ is skew-symmetric, it is a Lie bracket and
the triple $(D,[\cdot,\cdot]_D,\rho)$ defines a Lie algebroid.
When $M$ is a point $M=\mathrm{\{pt\}}$ and $\rho = 0$, the Leibniz
algebroid becomes a Leibniz algebra.
Given these definitions, we now introduce the Courant algebroids.
\begin{define}[Courant algebroid]
\label{def:Courant_algebroid}
Let $E \xrightarrow{\pi} M$ be a vector bundle over a manifold $M$.
A \textit{Courant algebroid} is a quadruple $(E, [\cdot, \cdot]_D, \rho, (\cdot,
 \cdot))$ where $[\cdot, \cdot]_D$ is a bilinear bracket
on $\Gamma (E)$, $\rho : E \to TM$ is 
an anchor map, and $(\cdot,\cdot)$ is a non-degenerate bilinear form on $\Gamma (E)$.
They satisfy the following axioms for any $e_i \in \Gamma (E)$ and $f \in C^{\infty}(M)$:
\begin{enumerate}
\item The bracket $[\cdot, \cdot]_D$ satisfies the Leibniz identity \eqref{eq:Leibniz_identity}.
\item $\rho ([e_1,e_2]_D) = [\rho(e_1) , \rho(e_2)]_{TM}$.
\item $[e_1, f e_2]_D = f [e_1,e_2]_D + (\rho (e_1) \cdot f) e_2 $.
\item $[e,e]_D = \frac{1}{2} \mathcal{D} (e,e)$.
\item $\rho (e_1) \cdot (e_2,e_3) = ([e_1,e_2]_D, e_3) + (e_2, [e_1,e_3]_D)$.
\end{enumerate}
Here $\mathcal{D}$ is a generalized exterior derivative on $\Gamma (E)$.
\end{define}
The axioms 1,2,3 are just the ones for the Leibniz algebroid.
Therefore any Courant algebroids are Leibniz algebroids.
An alternative but equivalent definition based on a
skew-symmetric -- the Courant -- bracket is known \cite{Liu:1997}, 
which is anti-symmetrization of the Dorfman bracket.
The Courant algebroids play important roles in generalized geometry \cite{Gualtieri}.
This inherits T-duality structure in its Drinfel'd double.

Next we introduce the Vaisman algebroids.
\begin{define}[Vaisman algebroid]
\label{def:Vaisman_algebroid}
Let $E \xrightarrow{\pi} N$ be a vector bundle over a manifold $N$.
A \textit{Vaisman algebroid} is a quadruple $(E, \vbracket{\cdot}{\cdot}_D, \rho, (\cdot,
 \cdot))$ where $\vbracket{\cdot}{\cdot}_D$ is a bracket on $\Gamma
 (E)$, $\rho : E \to TN$ is 
an anchor map, and $(\cdot,\cdot)$ is a non-degenerate bilinear form on $\Gamma (E)$.
They satisfy the following axioms for any $e_i \in \Gamma (E)$ and $f
 \in C^{\infty} (M)$:
\begin{enumerate}
\item $\vbracket{e_1}{f e_2}_D = f \vbracket{e_1}{e_2}_D + (\rho (e_1) \cdot f) e_2 $.
\item $\rho (e_1) \cdot (e_2,e_3) = (\vbracket{e_1}{e_2}_D,e_3) + (e_2, \vbracket{e_1}{e_3}_D)$
\end{enumerate}
\end{define}
The bracket $\vbracket{\cdot}{\cdot}_D$ is called a D-bracket.
One can find an alternative definition based on a skew-symmetric bracket
which satisfies the same axioms above.
The skew symmetrization of a D-bracket is called a C-bracket.
Obviously, any Courant algebroids are Vaisman algebroids.
From the viewpoint of DFT, a Vaisman algebroid appears on the tangent
bundle of a $2D$-dimensional para-Hermitian manifold
\cite{Vaisman:2012ke, Vaisman:2012px, Svoboda:2018rci}.
In this case, the bracket is given by the D-bracket in DFT.
A remarkable property of the Vaisman algebroid is that its defining
bracket is composed of a double of Lie algebroids \cite{Mori:2019slw}. 
The skew-symmetric bracket of the Vaisman algebroid on a para-Hermitian
manifold is nothing but the C-bracket that governs the gauge symmetry of DFT.
One can switch the C- and D-brackets by the standard procedures \cite{Roytenberg:1999}. 
In order that the algebra of the gauge transformation of DFT closes, a
constraint, known as the strong constraint, should be imposed on all the
fields and gauge parameters \cite{Hull:2009zb}.
In this case, the D-bracket reduces to the Dorfman bracket of
generalized geometry and the Vaisman algebroid reduces to the Courant
algebroid.

In the next section, we introduce racks and rackoids which are
integrations of the Leibniz algebras and algebroids.

\section{Racks and rackoids} \label{sct:rackoids}
In this section, we introduce the notion of racks and rackoids.
The rack has been proposed as a global group-like object associated with its
infinitesimal counterpart -- the Leibniz algebra \cite{Kinyon:2004,
Covez:2010, Bordemann:2016}.
This observation can be generalized to those for Leibniz algebroids.
The corresponding global, groupoid-like structure is known as rackoids.
In the following, we first define racks and then generalize the notion
to rackoids.
This section is based on \cite{Laurent-Gengoux:2015}. More details can
be found there.
\subsection{Racks}
The definition of racks is the following.
\begin{define}[Rack]
\label{def:rack}
The set $S$ together with a binary operation 
$(x,y) \mapsto x \rack y$ for any $x, y \in S$
is called a \textit{rack} if the map $y \mapsto x \rack y$ is bijective and
the operation $\rack$ satisfies the following left \textit{self-distributivity}:
\begin{align}
x \rack (y \rack z) = (x \rack y) \rack (x \rack z),
\label{eq:sd}
\end{align}
for any $x, y, z \in S$.
$x \rack y$ and the map $y \mapsto x \rack y$ are called the rack product and
 the rack action of $x$ on $y$, respectively.
\end{define}
We note that in general, the rack product $\rack$ is not associative 
$x \rack (y \rack z) \not= (x \rack y) \rack z$.
Since the rack action $x \rack \cdot$ is a bijection, there exists the unique element $y \in S$ such that $x \rack y = z$ for any $z \in S$. 
This implies the left invertibility of the rack product $\rack$.
A right rack is defined similarly for a right rack product $\triangleleft$.
In the following, we employ left racks and never consider the right ones.

An example of the rack product is the conjugation $g \rack h = g h
g^{-1}$ on a group $G$. 
This example trivially satisfies the self-distributivity.
In this sense, racks are defined over wrack of group structures.
There remains only the conjugation operation out of the group multiplication. 

There does not necessarily exist unit element in racks. We introduce pointed racks which are racks with the unit element $1$ with respect to the rack product.
\begin{define}[Pointed rack]
\label{def:pointed_rack}
Let $(S,\rack)$ be a rack.
When there is an element $1 \in S$ such that the following relation
 holds for any $x \in S$, 
\begin{align}
1 \rack x = x, \quad x \rack 1 = 1,
\end{align}
then $(S,\rack)$ is called a \textit{pointed rack}.
\end{define}
One can show that a group $G$ with the rack product defined by the
conjugation $g \rack h = g h g^{-1}$ is an example of a pointed rack
$(G,\rack)$. The element $1$ is obviously the unit element of the group $1 \in G$.
We next define a Lie rack.
\begin{define}[Lie rack]
\label{def:Lie_rack}
Let $S$ be a manifold.
When all the structures in a pointed rack $(S,\rack)$ are smooth and the
 rack action $y \mapsto x \rack y$ for any $x,y \in S$ is a
 diffeomorphism, then $(S,\rack)$ is called a \textit{Lie rack}.
\end{define}
We can then consider a tangent space of $(S,\rack)$ at the unit element.
Indeed, it is shown that an infinitesimal algebra defined on the tangent
space at a unit element of a Lie rack is nothing but a Leibniz algebra \cite{Kinyon:2004}.
This is quite analogous to the fact that the algebra on the tangent
space at the unit element of a Lie group $G$ is isomorphic to the Lie
algebra $\mathfrak{g}$ of $G$.

\subsection{Rackoids}
We next generalize the notion of the rack to that of rackoids.
One finds that this leads to an integration of the Leibniz algebroids.
The notion of rackoids has been introduced as a base ground of an integration
of the Courant algebroids \cite{Laurent-Gengoux:2015}.
Before going to the rackoids, let us begin with precategories.
\begin{define}[Precategory]
\label{def:small_precategory}
Let $({\cal G},M)$ be a pair composed of sets ${\cal G}$ and $M$.
When there exist surjection maps $s,t : {\cal G} \to M$, then $({\cal G},M)$ is called a semi-precategory. Here $s,t$ are called the source and the target maps.
When there is a unit map $\epsilon: M \to {\cal G}$ satisfying $s \circ \epsilon = t \circ
 \epsilon = \mathrm{id}_M$, then $({\cal G},M)$ is called a \textit{precategory}.
For each $x \in M$, we express $\epsilon (x) = 1_x \in {\cal G}$.
\end{define}
An element $g$ of ${\cal G}$ is regarded as a morphism from $s(g)$ to $t(g)$.
In the following, ${\cal G} \cat M$ denote a precategory.
We also note that $\epsilon (x)$ satisfying $s \circ \epsilon = t \circ \epsilon
= \mathrm{id}_M$ corresponds to a unit morphism at $x \in M$ if there is 
a composition map of morphisms.
When there exists an associative composition of morphisms, then ${\cal G} \cat M$ becomes a category.
In addition, when there exists an inverse for all the morphisms, ${\cal G} \cat M$ is a groupoid.

Since there are not necessarily compositions of morphisms in
(semi-)precategories, we next define bisections which enable us to find
an appropriate action of 
${\cal G}$ on $M$.
\begin{define}[Bisection]
\label{def:bisection}
Let ${\cal G} \rightrightarrows M$ be a semi-precategory.
A \textit{bisection} of ${\cal G}$ is defined by the following equivalent data:
\begin{enumerate}
\item A subset $\Sigma \subset {\cal G}$ such that the restricted source and
      the target maps $s,t : \Sigma \to M$ are bijection.
\item A map $\underline{\sigma} = t \circ \sigma: M \to M$ that is
      bijection. Here $\sigma : M \to \Sigma$ is a right inverse of $s$,
      namely, it is defined by $s \circ \sigma = \mathrm{id}_M$.
      
\end{enumerate}
\end{define}
In the following, maps associated with bisections $\Sigma, T, \cdots$
are denoted by $\sigma, \tau, \cdots$.
Now, a set of all the morphisms $g \in {\cal G}$ that satisfy $s(g) = x,
t (g) = y$ for all $x,y \in M$ is denoted by ${\cal G}_x^y$.
Then rackoids are defined as follows.
\begin{define}[Rackoid]
\label{def:rackoids}
For a semi-precategory ${\cal G} \rightrightarrows M$, 
a bisection $\Sigma \subset {\cal G}$ and $g \in {\cal G}^y_x$, one defines 
an action of $\Sigma$ on $g$
\begin{align}
\rack : (\Sigma, g) \mapsto \Sigma \rack g \in {\cal G}^{\underline{\sigma}
 (y)}_{\underline{\sigma} (x)}.
\end{align}
For bisections $\Sigma, T \subset {\cal G}$, we define $\Sigma \rack T$ as the
 image of an assignment $\Sigma \rack (\cdot)$ on $T$.
When the action $\rack$ satisfies the following properties,
\begin{enumerate}
\item For any bisections $\Sigma$, an assignment $\Sigma \rack (\cdot) :
       {\cal G} \to {\cal G}$ is bijective.
\item For any bisections $\Sigma, T$ and any $g \in {\cal G}$, the action
      $\rack$ satisfies the following self-distributivity,
\begin{align}
\Sigma \rack (T \rack g) = (\Sigma \rack T) \rack (\Sigma \rack g).
\end{align}
\end{enumerate}
then, this becomes a rack action and $({\cal G} \rightrightarrows M, \rack)$ is
 called a non-unital \textit{rackoid}.
In addition, for any $x \in M$, $g \in {\cal G}$, when there exists $\epsilon (x) = 1_x \in {\cal G}$
  such that 
 \begin{align}
 1_M \rack g = g, \quad \sigma(x) \rack 1_x = 1_{\underline{\sigma} (x)},
 \end{align}
then $({\cal G} \rightrightarrows M, \rack)$ is called a unital (or pointed)
  rackoid.
 Here $1_M$ stands for the bisection $\epsilon (M)$, namely, 
 the collection of $1_x$ for all $x \in M$.
When all the structures defined above are smooth, $({\cal G} \cat M, \rack)$ is
 called a Lie rackoid.
\end{define}
Note that since the map is bijective, the image of the map $\Sigma \rack
(\cdot)$ on a bisection $T$ is a bisection.
The geometrical meaning of the rack product defined above is obvious.
The rack action by $\Sigma$ shifts the initial (source) and the end (target)
points of $g \in {\cal G}$ along the $\Sigma$-direction. 
This is understood by the following relation of the associated
diffeomorphism on $M$:
\begin{align}
\underline{\sigma \rack \tau} = \underline{\sigma} \circ \underline{\tau} \circ \underline{\sigma}^{-1}.
\end{align}
Therefore the rack action is essentially a conjugation.
The rackoid defined in this way is a groupoid-like generalization of the rack.
Indeed, when $M$ is a point $M = \{\mathrm{pt}\}$, namely, there is only
one point in $M$, then the rackoid becomes a rack. This corresponds to the fact that a groupoid over a point $M =
\{\mathrm{pt}\}$ becomes a group.
In the following, all the collections of bisections in a smooth
precategory ${\cal G} \rightrightarrows M$ is denoted by $\mathrm{Bis}({\cal G})$.
Since all the structures in a Lie rackoid is smooth, we can now discuss
its infinitesimal algebra on the tangent space.
Parallel to the fact that a Lie algebroid is an infinitesimal object of a Lie groupoid,
one can define a Leibniz algebroid as an infinitesimal counterpart of a
Lie rackoid. Let us sketch this procedure in the following.

Let ${\cal G} \cat M$ be a unital (pointed) Lie rackoid.
Through the unit map $\epsilon : M \hookrightarrow {\cal G}$, we identify $M$
to a subset of ${\cal G}$. Namely, $x \in M$ is identified with $1_x \in {\cal G}$.
Fibers of the pullback bundle $\epsilon^*T {\cal G}
 \xrightarrow{\pi^*} M$ by $\epsilon$ are given by $T_{1_x} {\cal G}$.
We consider differentials of source and the target maps of ${\cal G} \cat M$, 
$Ts,Tt : T{\cal G} \to TM$.
By definition, since the fibers of the pullback bundle $\epsilon^*T{\cal G}$ are
just copies of the $T{\cal G}$ fibers, diffeomorphisms $\epsilon^*T{\cal G} \to  TM$ are naturally defined. 
Explicitly, we have the induced maps,
\begin{align}
T_{1_x} t: T_{1_x} {\cal G} \to T_x M,  \qquad 
T_{1_x} s: T_{1_x} {\cal G} \to T_x M.
\end{align}
Using two maps, we define an infinitesimal algebroid $\mathcal{A}$ of
a unital Lie rackoid.
\begin{define}[Infinitesimal algebroids of rackoid]
\label{def:infinitesimal_algebroid} 
Given a unital (pointed) Lie rackoid ${\cal G} \cat M$, 
we have a vector bundle $\mathcal{A}$ over $M$ defined by 
\begin{align}
\mathcal{A} = \mathrm{Ker} (Ts) = \coprod_{x \in M}
 \mathrm{Ker} (T_{1_x} s) \in \coprod_{x \in M} T_{1_x} {\cal G}.
\end{align}
This is called an infinitesimal algebroid of ${\cal G}$.
The anchor $\rho : \mathcal{A} \to TM$ is defined by 
\begin{align}
\rho = - Tt|_{\mathcal{A}}.
\end{align}
\end{define}
Since all the structures discussed above are well-defined, we consider 
the adjoint action on $\mathcal{A}$ induced by the rack action
$\Sigma^{\rack}$ of a bisection $\Sigma$:
\begin{align}
\mathrm{Ad}_{\Sigma} = T_{1_x} \Sigma^{\rack}|_{\mathcal{A}} :
 \mathcal{A}_x \to \mathcal{A}_{\underline{\sigma} (x)},
\end{align}
The rack action satisfies, by definition, the self-distributivity.
Namely, for any bisections $\Sigma, T \subset {\cal G}$ and $g \in {\cal G}$, we have
$\Sigma^{\rack} \circ T^{\rack} = (\Sigma \rack T)^{\rack} \circ \Sigma^{\rack}$.
Since $T_{1_x} \Sigma^{\rack}$ induces the adjoint action, 
the self-distributive relation of the rack action results in the
composition of the adjoint action:
\begin{align}
\mathrm{Ad}_{\Sigma} \circ \mathrm{Ad}_T = \mathrm{Ad}_{\Sigma \rack T}
 \circ \mathrm{Ad}_{\Sigma}.
\label{eq:ad_sd}
\end{align}
The following lemma enable us to make contact with explicit derivation
of the bracket in the algebroid $\mathcal{A}$:
\begin{lemma}[Family of bisections]
\label{lem:bisection_family}
Let $({\cal G} \cat M, \rack)$ and $\mathcal{A}$ be a Lie rackoid and its infinitesimal algebroid.
Let us consider the section $\Gamma : M \to \epsilon^* T{\cal G}$.
Namely, $\Gamma : x \mapsto b_x \in \mathcal{A}_x \subset T_{1_x} {\cal G}$.
Then, there is a family of bisection, $(\Sigma_u)_{u \in I}$, $I =
 (-1,1)$ of ${\cal G}$ such that $\Sigma_0 = \epsilon (M)$ and $\frac{\partial}{\partial u}
 \sigma_u (x) |_{u=0}$ coincides with $b_x$ for all $x \in M$.
Here $\sigma_u$ is a map $\sigma_u : M \to {\cal G}$ associated with the 
bisection $\Sigma_u$ and it is parametrized by $u$.
This satisfies $s \circ \sigma_u = \mathrm{id}_M$, $\sigma_u (x) \in \Sigma_u
 \subset {\cal G}$.
In this case, the assignment $x \mapsto \frac{\partial}{\partial u} \sigma_u
 (x)|_{u=0}$ is a smooth section of $\mathcal{A}$.
Since $\Sigma_{u=0} = \Sigma_0 = \epsilon (M) = \coprod_{x \in M} 1_x$, we have 
 $\sigma_{u=0} (x) = 1_x$.
\end{lemma}
The proof is found in \cite{Laurent-Gengoux:2015}.
With this fact at hand, we find that the tangent space of $\text{Bis}({\cal G})$ at 
$g = \epsilon (M)$ is $\Gamma (\mathcal{A})$.
Using the adjoint map for families of bisections, we now define the bracket structure on $\Gamma
(\mathcal{A})$.
\begin{define}[Bracket in $\mathcal{A}$]
\label{def:bracket_A} 
For any $a,b \in \Gamma (\mathcal{A})$, a bracket $[\cdot,\cdot] :
 (b,a) \mapsto [b,a]$ is defined by
\begin{align}
[b,a] = \frac{\partial}{\partial u} \mathrm{Ad}_{\Sigma_u} a|_{u=0}.
\label{eq:A_bracket}
\end{align}
Here $\Sigma_0 = \epsilon (M)$ and $\frac{\partial}{\partial u} \sigma_u
 (x)|_{u=0} = b_x$. $\Sigma_u, u \in (-1,1)$ is a family of smooth
 bisection of ${\cal G}$.
\end{define}
One can check the bracket $[\cdot,\cdot]$ defined above satisfies the Leibniz
identity \eqref{eq:Leibniz_identity} as follows.
By the self-distributivity of the rack product for families of
bisections $\Sigma_u, T_v$ acting on any $a \in \Gamma (\mathcal{A})$, we have
\begin{align}
\mathrm{Ad}_{\Sigma_u} \circ \mathrm{Ad}_{T_v} a = \mathrm{Ad}_{\Sigma_u
 \rack T_v} \circ \mathrm{Ad}_{\Sigma_u} a.
\end{align}
By differentiating both sides with respect to $v$ and evaluating at $v=0$
, we find
\begin{align}
\mathrm{Ad}_{\Sigma_u} \circ [b,a]
 = 
\frac{\partial}{\partial v} \mathrm{Ad}_{\Sigma_u \rack T_v} |_{v=0}
 \circ 
\mathrm{Ad}_{\Sigma_u} a.
\label{eq:sd_diff}
\end{align}
Here we have used the relation $[b,a] = \frac{\partial}{\partial v} \mathrm{Ad}_{T_v} a |_{v=0}$, 
$b_x = \frac{\partial}{\partial v} \tau_v (x) |_{u=0}$.
Since the rack action on $\Gamma (\mathcal{A})$ induces the adjoint
action, we have
\begin{align}
\frac{\partial}{\partial v} \Sigma_v \rack \tau_u (x) |_{v=0} =& \ 
\Sigma_u \rack \frac{\partial}{\partial v} \tau_v (x) |_{v=0}
\notag \\
=& \ \mathrm{Ad}_{\Sigma_u} \, b_x.
\end{align}
Then the right hand side of \eqref{eq:sd_diff} becomes $[\mathrm{Ad}_{\Sigma_u} b, \mathrm{Ad}_{\Sigma_u} a]$.
By differentiating both sides again with respect to $u$ and setting $u=0$, we obtain 
\begin{align}
\frac{\partial}{\partial u} \mathrm{Ad}_{\Sigma_u} ([b,a]) |_{u=0}
=& \ 
[ 
\frac{\partial}{\partial u} \mathrm{Ad}_{\Sigma_u} b|_{u=0}
,
\mathrm{Ad}_{\Sigma_u} a|_{u=0}
]
+
[
\mathrm{Ad}_{\Sigma_u}b |_{u=0}
,
\frac{\partial}{\partial u} \mathrm{Ad}_{\Sigma_u} a|_{u=0}
].
\end{align}
The left hand side gives $[c,[b,a]]$ while the first and the second terms in the right
hand side are $[[c,b],a]$ and $[b, [c,a]]$. Here we have used the facts
$\frac{\partial}{\partial u} \mathrm{Ad}_{\Sigma_v} b|_{u = 0} = [c,b]$,
$\mathrm{Ad}_{\Sigma_u} a |_{u=0} = a$.
Then the Leibniz identity follows:
\begin{align}
[c,[b,a]] = [[c,b],a] + [b,[c,a]].
\label{eq:bracket_Leibniz_identity}
\end{align}
One also finds that due to the definition of the anchor map $\rho = - T
t|_{\mathcal{A}}$, the Leibniz rule for the bracket holds.
We then end up with the following theorem.
\begin{theorem}[Tangent Leibniz algebroid of Lie rackoid]
\label{def:tangent_Leibniz_algebroid} 
Let $({\cal G} \cat M, \rack)$ be a unital Lie rackoid over a manifold $M$.
There is a vector bundle $\mathcal{A} = \mathrm{Ker} (Ts) \to M$ in the
 induced bundle $\epsilon^* T{\cal G}$. An anchor $\rho = - Tt|_{\mathcal{A}}$
 and a bracket $[b,a] = \frac{\partial}{\partial u}
 \mathrm{Ad}_{\Sigma_u} a|_{u=0}$ is defined for a family of bisections $\Sigma_u$.
Then the triple $(\mathcal{A}, [\cdot, \cdot], \rho)$ defines a Leibniz
 algebroid over $M$. This is called a \textit{tangent Leibniz algebroid} of ${\cal G}$.
\end{theorem}
We stress that the Leibniz identity \eqref{eq:bracket_Leibniz_identity}
derived by the adjoint map \eqref{eq:ad_sd} originates from the self-distributive
relation of the rack product.
This fact will be an important clue for the integration of the Vaisman
algebroid.

\section{Pre-rackoids and doubled cotangent paths} \label{sct:pre-rackoids}
In this section, we exhibit an explicit example of the Lie rackoid through the cotangent paths
discussed in \cite{Laurent-Gengoux:2018zoz}. 
This is a basic ground for the group-like, global structures 
of the standard Courant algebroid.
We first introduce the standard Courant algebroid and discuss the rackoid structure associated with it. 
We then proceed to the rackoid-like structure for the Vaisman algebroid.
A key ingredient is the notion of the pre-rackoid based on the doubled
foliations of a para-Hermitian manifold.

\subsection{Cotangent path rackoids and Courant algebroid}
The most familiar example of the Courant algebroid is the standard Courant
algebroid. The vector bundle of the standard Courant algebroid is given
by the generalized tangent bundle $\mathbb{T}M = TM \oplus T^*M$ over a
manifold $M$.
The Dorfman bracket of the standard Courant algebroid is given by
\begin{align}
[e_1, e_2]_D = [X_1, X_2]_{TM} + \mathcal{L}_{X_1} \xi_2 -
 \iota_{X_2} \mathrm{d} \xi_1.
\label{eq:Dorfman_bracket_standard}
\end{align}
Here $e_i = X_i + \xi_i, \ (i=1,2)$ and $X_i \in \Gamma (TM)$ are vector fields,
$\xi_i \in \Gamma (T^*M)$ are 1-forms. The bracket $[\cdot, \cdot]_{TM}$ in
the right hand side is the ordinary Lie bracket of the vector fields, $\mathcal{L}_{X}$, $\iota_X$ are the Lie
derivative and the interior product associated with $X$ and $\mathrm{d}$
is the exterior derivative operator.
In \cite{Laurent-Gengoux:2018zoz}, a rackoid $({\cal G} \cat M, \rack)$
that results in a tangent Leibniz algebroid equipped with the bracket
\eqref{eq:Dorfman_bracket_standard} is constructed.
Here ${\cal G} = P T^*M$ is a set of paths $P T^*M = C^{\infty} ([0,1], T^*M)$ on a compact
manifold $M$ and the rack product $\rack$ is defined by automorphisms on
the Dorfman bracket.

In the following, we briefly introduce the discussions in \cite{Laurent-Gengoux:2018zoz}
and then generalize the construction to the one in the doubled geometry.
Before examining the space $PT^*M$, we first clarify a precategory structure for paths on $M$.
\begin{define}[Precategory by paths]
\label{def:precategory_path}
Let $M$ be a compact manifold of finite dimensions.
Let $PM$ be a set of smooth paths $\gamma : [0,1] \to M$.
The source and the target maps are defined by 
\begin{align}
s (\gamma) = \gamma_0, \quad t (\gamma) = \gamma_1,
\end{align}
where the path $\gamma_t$ is parametrized by $t \in [0,1]$.
The unit map $\epsilon : M \to PM$ is defined by a map to a constant $c$, $\epsilon (x) = c$ for any $x \in M$. 
Then $PM \cat M$ becomes a smooth precategory.
\end{define}
Since the path can be seen as a map $\gamma : [0,1] \to C^{\infty}
(M,M)$, bisections of the precategory $PM \cat M$ is defined as follows.
\begin{define}[$\text{Bis}(PM)$ of $PM \cat M$]
\label{def:bisection_path}
A set of smooth bisections $\text{Bis}(PM)$ in the precategory $PM \cat
 M$ is defined by
\begin{align}
\text{Bis} (PM) = 
\{
\gamma_0 = \mathrm{id}_M , \ \gamma_1 \ \text{is a diffeomorphism
 of } M
\}.
\end{align}
\end{define}
Namely, there is one to one correspondence between a pair of fixed
points $x = \gamma_0 (x)$, $y = \gamma_1 (x)$ and a path in a bisection of $PM \cat M$.
We then define the path rackoids.
\begin{define}[Path rackoid]
\label{def:path_rackoid}
Given a precategory $PM \cat M$ defined by paths $\gamma : [0,1] \to
 M$, we introduce the rack product $\rack$ for any elements of bisections $\psi,
 \varphi$ in $\text{Bis}(PM)$ and for any $t \in [0,1]$ as 
\begin{align}
(\psi \rack \varphi)_{t} = \psi_1 \circ \varphi_t \circ \psi_1^{-1}.
\label{eq:path_rackoid_product}
\end{align}
Further, we define the rack action of a bisection $\psi$ on a path
 $\gamma \in PM$ as 
\begin{align}
(\psi \rack \gamma)_t = \psi_1\circ \gamma_t.
\label{eq:path_rackoid_product2}
\end{align}
Then it is obvious that the product $\rack$ satisfies the
 self-distributivity and $(PM \cat M, \rack)$ becomes an infinite-dimensional unital Lie rackoid.
We call this a \textit{path rackoid}.
\end{define}
As we have discussed in the previous section, the rack action of $\psi$ on
$\phi$ shifts the initial and the end points of $\phi$ along the $\psi$
direction. 
Since the path in the base space $M$ induces the path in the cotangent
bundle $T^*M$, the notion of the path rackoid is generalized to that of
the cotangent path.
\begin{define}[Precategory by cotangent paths]
\label{def:cotangent_path}
Let $M$ be a finite-dimensional compact manifold and 
$T^*M \xrightarrow{\pi} M$ be the cotangent bundle over $M$.
We define a pair of smooth paths on $M$ as 
\begin{align}
PT^*M =
\left\{
(\gamma, \eta) : [0,1] \to T^*M \, | \, \eta : \text{smooth}
\right\}.
\end{align}
Here $\eta$ is the actual cotangent path on $T^*M$ and it is related to
 the path $\gamma$ in the base space $M$ through the projection $\gamma = \pi \circ
 \eta$.
The path $\eta$ is recognized as a morphism whose source and target maps
 $s,t$ are defined by $s = \pi \circ \mathrm{ev}_0$, $t = \pi \circ \mathrm{ev}_1$.
Here $\mathrm{ev}_{0,1}$ is the evaluation of $\eta$ at $t=0,1$ and takes values in $T^*M$. 
One notices that the projection results in the initial and the end points of
the path $\gamma$ on $M$.
Again, the unit map $\epsilon$ is defined by the constant path.
These structures make $PT^*M \cat M$ be a smooth precategory.
\end{define}

We next define bisections of the precategory $PT^*M \cat M$.
\begin{define}[$\text{Bis}(PT^*M)$ of $PT^*M \cat M$]
\label{def:bisection_cotangent_path}
Let $PT^*M \cat M$ be the precategory defined by the cotangent paths.
Bisections of $PT^*M \cat M$ are pairs of paths $\Sigma =
 (\phi,\eta) \subset PT^*M$ where each path is defined as follows.
First, the paths $\phi$ in the base space $M$ are bisections of the
 precategory $PM \cat M$.
Second, the cotangent paths $\eta$ are the section of the pullback bundle of $\phi_t : M \to M$ defined by $[0,1] \ni t
 \mapsto \phi_t^{*} \eta_t$. Namely, they are 1-forms on $M$.
Explicitly, the path of the 1-form is defined by 
\begin{align}
t \in [0,1] \mapsto \phi_t^* \eta_t (X_x) = \eta_{t} (T_x \phi_t (X_x))
 \quad \text{for} \ X_x \in T_x M, \ x \in M.
\end{align}
\end{define}
Given these definitions, we introduce a rack structure in bisections $\Sigma$ of the precategory $PT^*M \cat M$.
\begin{define}[Rack action of cotangent paths]
\label{def:rack_action_cotangent_path}
Let $PT^*M \cat M$ be a precategory defined by the cotangent paths.
For any bisections $\Sigma = (\phi, \eta)$, $T = (\psi, \zeta)$ of 
$PT^*M \cat M$, the rack action is defined by
\begin{align}
(\Sigma \rack T)_t = (\phi, \eta) \rack (\psi, \zeta)_t
=
\left(
\phi_1 \circ \psi_t \circ \phi_1^{-1}, \,
(\phi_1^{*})^{-1}
\left(
\zeta_t - \iota_{\dot{\psi}_t} \circ \phi_1^* \mathrm{d} \beta_{\Sigma}
\right)
\right),
\label{eq:cotangent_path_rack_product}
\end{align}
where $\dot{\psi}_t = \frac{d}{dt} \psi_t$ and 
\begin{align}
\beta_{\Sigma} = \int^1_0 \! ds \ \phi_s^* \eta_s
\end{align}
is a 1-form associated with $\Sigma$.
Similarly, for a cotangent path $a \in PT^*M$ composed of a cotangent vector $\theta_t
 (x), x \in M$ at $\gamma_t (x)$, where the path $[0,1] \ni t \mapsto
 \gamma_t$ is given by the projection $\gamma = \pi \circ a$, 
we define a rack action of a bisection $\Sigma = (\phi,\eta) \subset PT^*M$ on the
 cotangent path $a = (\gamma,\theta)$ as
\begin{align}
(\Sigma \rack a)_t (x) = 
\left(
\phi_1 \circ \gamma_t (x), \,
(\phi_1^{-1})^{*} \circ 
\left\{
\theta_t (x) - \iota_{\dot{\gamma}_t} \circ 
\phi_1^{*} \circ \mathrm{d} 
\beta_{\Sigma} (x)
\right\}
\right).
\label{eq:cotangent_path_rack_product2}
\end{align}
\end{define}
Note that for a path $(\gamma, \theta) \subset PT^*M$, 
$(\dot{\gamma},\theta)$ is a pair of a tangent vector and a 1-form and 
$(\dot{\gamma}_t, \theta_t)$ is regarded as $C^{\infty}
([0,1], \mathbb{T}M)$. Here $\mathbb{T}M = TM \oplus T^*M
\xrightarrow{\pi} M$ is the generalized tangent bundle over $M$.
The rack product by $\Sigma = (\phi, \eta)$ shifts the initial and the end
points of the path on the base space $M$.
On the other hand, the rack product in
\eqref{eq:cotangent_path_rack_product}, \eqref{eq:cotangent_path_rack_product2} induces the pull-back by $\phi$ and 
the gauge transformation by $\eta$ in the cotangent space.
It is easy to check that the first component of the rack action
\eqref{eq:cotangent_path_rack_product} satisfies the
self-distributivity. 
On the other hand, the self-distributivity of the rack action in the
1-forms is little bit non-trivial.
A careful analysis revealed that the second component of the cotangent
path also satisfies the self-distributivity.
One finds the detailed proof in \cite{Laurent-Gengoux:2018zoz}.
The proof is based on the explicit form of the 
exterior derivative of the 1-form associated with
the bisection $\Sigma \rack T$.
This is given by the direct calculations. The result is 
\begin{align}
\mathrm{d} \beta_{\Sigma \rack T} = 
(\phi_1^{*})^{-1} \mathrm{d} \beta_T - (\phi_1^{*})^{-1} \psi_1^* \phi_1^* \mathrm{d} \beta_{\Sigma} + \mathrm{d} \beta_{\Sigma}.
\label{eq:d_1-form}
\end{align}
Here bisections are expressed as $\Sigma = (\phi,\eta)$, $T = (\psi,
\zeta)$. 
This is a key expression to prove the self-distributivity of the rack product \cite{Laurent-Gengoux:2018zoz}.
We will see in the next subsection that this specific expression, 
that is necessary to prove the self-distributivity, 
 does not hold for the pre-rackoid.

Since the structure discussed above are all smooth 
and the rack action is well-defined, $(PT^*M \cat M, \rack)$ becomes a unital Lie
rackoid.
Now we are in a position to discuss the infinitesimal algebroid
$\mathcal{A}$ of the rackoid $(PT^*M \cat M, \rack)$.
Consider the $s^{-1}$-fiber over $M$.
The infinitesimal algebroid $\mathcal{A} = \mathrm{Ker} Ts \xrightarrow{\pi} M$
for the Lie rackoid $PT^*M \cat M$ is well-defined.
Recall that for the bisections, we have a tangent bundle $T(PT^*M) =
\left\{\dot{\gamma} : [0,1] \to TM \oplus T^*M \right\}$. 
Then given by the section $\Gamma : M \to PT^*M$, we define the bracket on 
$\Gamma (A) = \Gamma_* ([0,1], \mathfrak{X} (M) \oplus \Omega^1 (M))$ 
as clarified in the general discussion.
From the definition of the rack product of $(PT^*M \cat M, \rack)$, 
the bracket is calculated as the adjoint action associated with the rack action for
families of bisections.
We define the following quantities,
\begin{align}
&
\frac{\partial}{\partial u} \phi^u_t |_{u=0} = X_{1,t},
\qquad 
\frac{\partial}{\partial u} \psi^u_t |_{u=0} = X_{2,t},
\notag \\
&
\frac{\partial}{\partial u} \beta_{\Sigma^u_t} |_{u=0} = \alpha_{1,t},
 \quad 
\frac{\partial}{\partial u} \zeta^u_t|_{u=0} = - \alpha_{2,t},
\label{differentialequations}
\end{align}
where $\phi^u, \psi^u$ are families of bisections parametrized by $u$, 
$X_{i}, \alpha_{i} \ (i=1,2)$ are vector fields and 1-forms
on $M$.
Then we find 
\begin{align}
&
\frac{\partial}{\partial u} \frac{\partial}{\partial v}
\phi_1^v \circ \psi_t^u \circ (\phi_1^v)^{-1} |_{u=v=0}
= [X_{1,t=1}, X_{2,t}], 
\notag \\
&
\frac{\partial}{\partial u} \frac{\partial}{\partial v} (\phi_1^u)^{-1*}
 (\zeta_t^v) |_{u = v = 0} = \mathcal{L}_{X_{1,t=1}} \alpha_{2,t}.
\end{align}
Therefore, the adjoint action associated with the rack action $\Sigma \rack (\cdot)$
results in the following bracket:
\begin{align}
[
X_{1,t} + \xi_{1,t}, X_{2,t} + \xi_{2,t}
]_D = 
[X_{1,t=1}, X_{2,t}]_{TM}
+
\mathcal{L}_{X_{1,t=1}} \alpha_{2,t} - \iota_{\dot{X}_{2,t}} \mathrm{d} \int^1_0 \! ds
 \ \alpha_{1,s}.
\label{eq:pre_Courant_bracket}
\end{align}
This by definition satisfies the left Leibniz identity.
The bracket is almost the Dorfman bracket of the standard Courant algebroid.
We then define the following subbundle of $\mathcal{A}$:
\begin{align}
I =  
\left\{
(X_t, \xi_t) | X_1 = 0, \
\int_0^1 \! ds \ \alpha_s = 0
\right\}
 \end{align}
This defines an isomorphism between $\mathcal{A}/I$ and the standard Courant algebroid via the map 
$\varphi = (\mathrm{ev}_1, \int^1_0 \! dt) : \mathcal{A} \to \mathbb{T}M$.
Indeed, we have 
\begin{align}
\varphi : (X_t, \alpha_t) \mapsto (X_1, \int_0^1 \! dt \ \alpha_t).
\end{align}
Using these facts, by defining $X_i = X_{i,t=1}$, $\xi_i = \int_0^1 \alpha_{i,s} ds$, we
finally obtain 
\begin{align}
\varphi 
[X_{1,t} + \alpha_{1,t}, X_{2,t} + \alpha_{2,t}]_D
= 
[X_1, X_2]_{TM}
+
\mathcal{L}_{X_1} \xi_2 - \iota_{X_2} \mathrm{d} \xi_1
.
\end{align}
This is nothing but the Dorfman bracket
\eqref{eq:Dorfman_bracket_standard} of the standard Courant algebroid.

\subsection{Pre-rackoids and doubled cotangent paths}
Exploiting the discussions in the previous subsections, we next explore the 
group-like, global structure associated with the Vaisman algebroid.
To this end, we look for a rackoid-like structure whose bracket in the infinitesimal algebroid results in the bracket of the Vaisman algebroids.
An explicit example of the Vaisman algebroid appears in the tangent
bundle of a para-Hermitian manifold $\mathcal{M}$ \cite{Vaisman:2012ke, Svoboda:2018rci}.
The bracket is given by the D-bracket:
\begin{align}
\vbracket{e_1}{e_2}_D =& \ 
[X_1, X_2]_{T\mathcal{M}_+}
+ \mathcal{L}_{\xi_1} X_2 - \iota_{\xi_2} \mathrm{d}^* X_1
\notag \\
+ & \ 
[\xi_1, \xi_2]_{T\mathcal{M}_-}
+ \mathcal{L}_{X_1} \xi_2 - \iota_{X_2} \mathrm{d} \xi_1.
\label{eq:D-bracket}
\end{align}
Here $e_i = X_i + \xi_i$ and $X_i$, $\xi_i$ are vectors and dual vectors, 
$\mathrm{d}^*, \mathrm{d}$ are exterior derivatives on the vector and its
dual vector spaces.
$[\cdot, \cdot]_{T\mathcal{M}_+}, [\cdot, \cdot]_{T\mathcal{M}_-}$ are
Lie brackets on vector and dual vector spaces and $\mathcal{L}_X,
\mathcal{L}_{\xi}$ are Lie derivatives associated with the vectors and
their duals.
There are two independent parts in the D-bracket.
The first term in the first line and the second, third terms in the
second line define the Dorfman bracket
\eqref{eq:Dorfman_bracket_standard} for the standard Courant algebroid. 
The other terms are necessary for the Vaisman algebroid.
As we have observed, brackets in the Vaisman algebroids do not necessarily satisfy the Leibniz identity.
Indeed, the Leibniz identity of the D-bracket \eqref{eq:D-bracket} never
holds without the strong constraint.
If one solves the strong constraint in DFT by the para-holomorphic quantities
in the para-Hermitian manifold, the latter all vanish \cite{Mori:2019slw}.

A skew-symmetric bracket is defined by the anti-symmetrization of the D-bracket.
\begin{align}
\vbracket{e_1}{e_2}_C =& \ 
[X_1, X_2]_{T\mathcal{M}_+} + \mathcal{L}_{\xi_1} X_2 - \mathcal{L}_{\xi_2} X_1 -
 \frac{1}{2} \mathrm{d} (\iota_{X_1} \xi_2 - \iota_{X_2} \xi_1)
\notag \\
+& \ [\xi_1, \xi_2]_{T\mathcal{M}_-} + \mathcal{L}_{X_1} \xi_2 - \mathcal{L}_{X_2}
 \xi_1 - \frac{1}{2} \mathrm{d}^* (\iota_{\xi_1} X_2 - \iota_{\xi_2} X_1)
\end{align}
This is nothing but the C-bracket which governs the gauge symmetry of DFT.

As we have seen in the previous sections, the origin of the Leibniz identity is the self-distributivity of the rack action.
Therefore it is natural that 
a global counterpart of the Vaisman algebroid is given by a rackoid without self-distributive rack actions.
To elucidate such a structure, we first define the notion of pre-rackoids.
\begin{define}[Pre-rackoid]
\label{def:pre-rackoid}
Let ${\cal G} \cat M$ be a semi-precategory.
Bisections of ${\cal G}$ are defined as in the definition in section \ref{sct:rackoids}.
For any bisection $\Sigma$ and $g \in {\cal G}^y_x$, we define an action of
 $\Sigma$ on $g$
\begin{align}
\prack : (\Sigma, g) \mapsto \Sigma \prack g \in
 {\cal G}^{\sigma \cdot y}_{\sigma \cdot x}.
\end{align}
Here $\sigma \cdot x$ stands for a smooth action of $\sigma$ on $x \in M$.
When the assignment $\Sigma \prack (\cdot): {\cal G} \to {\cal G}$ is bijective, we call this the pre-rack action (product).
We then call $({\cal G} \cat M, \prack)$ the pre-rackoid.
We can introduce an additional unital structure by $\epsilon$ by which we define the unital pre-rackoid.
\end{define}
Similar to the rack action, the pre-rack action of $\Sigma$ on $g \in {\cal G}$ 
is defined by the shift of the initial and the end points of the morphism $g$ by the $\Sigma$ action.
However, this action is not given by $\underline{\sigma} = t \circ \sigma$ in general.
Due to this property, we stress that $\prack$ does not necessarily satisfy the self-distributivity.
We will see the explicit example of this pre-rack action in the following.
We note that the smooth pre-rack action $\prack$ still induces a bracket
via the adjoint action.

In order to find an explicit example of the pre-rackoid, we once again write down the necessary conditions.
(i) The pre-rack product $\prack$ does not satisfy the
self-distributivity in general.
(ii) Under the imposition of the strong constraint in DFT, the pre-rack
product $\prack$ reduces to the rack product $\rack$ that satisfies the self-distributivity.
(iii) A bracket obtained by the induced adjoint map of the pre-rack action
$\prack$ is given by the D-bracket \eqref{eq:D-bracket}.

A key ingredient is the doubled structure of the D-bracket.
To incorporate this structure, we begin with a $2D$-dimensional para-Hermitian manifold $\mathcal{M}$ as the base space
of the pre-rackoid.
The local coordinates $x^M = (x^{\mu}, \tilde{x}_{\mu})$ of doubled
geometry in DFT naturally appears in a para-Hermitian manifold \cite{Freidel:2017yuv,Freidel:2018tkj}.
Due to the para-complex structure $K$ such that $K^2 = 1$ of $\mathcal{M}$, the tangent bundle $T\mathcal{M}$ is decomposed into two parts $T\mathcal{M} =
T\mathcal{M}_+ \oplus T\mathcal{M}_-$.
Here each part is determined by the eigenbundles of the para-complex structure $K^2=1$.
There are doubled foliations $\mathcal{F},\tilde{\mathcal{F}}$ of
$\mathcal{M}$ associated with the integrability of $T\mathcal{M}_+$ and $T\mathcal{M}_-$.
The leaves associated with $T\mathcal{M}_+$ are characterized by spaces where
$\tilde{x} = \text{const}$. 
We express a leaf defined by a locus for fixed $\tilde{x}$ by
$\mathcal{F}_{x,[\tilde{x}]}$. The coordinate along the space
$\mathcal{F}_{x,[\tilde{x}]}$ is $x^{\mu}$.
The same is true for $T\mathcal{M}_-$, namely, 
spaces defined by $x = \text{const}$ is denoted by $\tilde{\mathcal{F}}_{[x],\tilde{x}}$.
In this picture, solving the strong constraint and defining the physical space is
equivalent to choose a leaf in the foliations of $\mathcal{M}$.
For example, the strong constraint in DFT is trivially solved by the para-holomorphic quantities, 
{\it i.e.}, those that depend only on $x^{\mu}$ coordinates.
This implies that we choose a $\mathcal{F}_{x,[\tilde{x}]}$ space as a physical spacetime.

\begin{figure}[t]
\begin{center}
\includegraphics[scale=0.45]{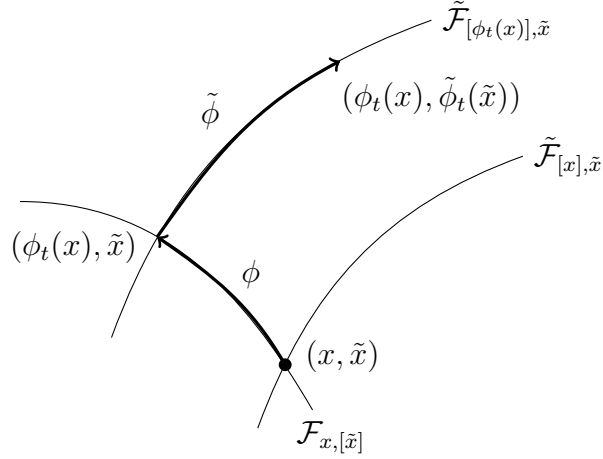}
 \caption{Leaves for the doubled foliations of $\mathcal{M}$ (thin lines). The doubled path is the
 concatenation of the paths along the leaves $\mathcal{F}_{x,[\tilde{x}]}$ and
 $\tilde{\mathcal{F}}_{[\phi_t (x)],\tilde{x}}$ (bold lines).}
 \label{fig:doubled_path}
\end{center}
\end{figure}

Let us consider a path $(\phi, \eta) \subset PT^*\mathcal{M} $ in a leaf $\mathcal{F}_{x,[\tilde{x}]}$ given by $\tilde{x} = \text{const}.$ 
As we have discussed in the previous subsection, one can define a cotangent path rackoid $(PT^*\mathcal{M}
\cat \mathcal{F}_{x,[\tilde{x}]}, \rack)$ whose base space is $\mathcal{F}_{x,[\tilde{x}]}$.
We also define another cotangent path rackoid $(\widetilde{PT^*\mathcal{M}} \cat
\tilde{\mathcal{F}}_{[x],\tilde{x}}, \rack)$ based on the cotangent path $(\tilde{\phi}, \tilde{\eta}) \subset \widetilde{PT^*\mathcal{M}}$ on the leaf $\tilde{\mathcal{F}}_{[x],\tilde{x}}$ at
$x = \text{const}$.
These rackoids $(PT^*\mathcal{M},\widetilde{PT^*\mathcal{M}})$ are defined independently.
Now we introduce a new path based on a pair of paths $(\phi, \eta) \subset PT^*\mathcal{M} $, $(\tilde{\phi}, \tilde{\eta}) \subset \widetilde{PT^*\mathcal{M}}
$ in the doubled foliations on $\mathcal{M}$ and the cotangent bundle $T^*\mathcal{M}$.
The new path on the base space $\mathcal{M}$ is defined by the concatenation of the paths
$\phi: [0,1] \to \mathcal{F}_{x,[\tilde{x}]}$ and $\tilde{\phi} : [0,1]
\to \tilde{\mathcal{F}}_{[\phi_t (x)],\tilde{x}}$ along the leaves
$\mathcal{F}_{x,[\tilde{x}]}$ and $\tilde{\mathcal{F}}_{\phi_t(x),[\tilde{x}]}$,
respectively (see Fig \ref{fig:doubled_path}). 
Here $t \in [0,1]$ is the parameter of the paths.
More explicitly, we have the path acting on the point $(x,\tilde{x}) \in
\mathcal{M}$ as 
\begin{align}
(\phi, \tilde{\phi})_t (x,\tilde{x}) = 
(\phi_t (x), \tilde{\phi}_{t} (\tilde{x}))
\end{align}
The paths in the cotangent space is defined
similarly by the concatenation of $\eta$ and $\tilde{\eta}$
on $(T\mathcal{M}_+)^*$ and $(T\mathcal{M}_-)^*$.
We call this the doubled cotangent path and denote it $PT^*\mathcal{M} \diamond
\widetilde{PT^*\mathcal{M}} \equiv \mathbf{PT}^*\mathcal{M}$.
We define the source and the target maps $\mathbf{PT}^*\mathcal{M} \to \mathcal{M}$
as $s$ and $\tilde{t}$.
Here $s$, $\tilde{t}$ are the source and the target maps of $PT^*\mathcal{M} \cat
\mathcal{F}_{x,[\tilde{x}]}$ and $\widetilde{PT^*\mathcal{M}} \cat \tilde{\mathcal{F}}_{[\phi_t(x)],\tilde{x}}$.
Then, $\mathbf{PT}^*\mathcal{M} \cat \mathcal{M}$ becomes a semi-precategory.
If we employ the pair of the unit maps $(\epsilon, \tilde{\epsilon})$ of 
$PT^*\mathcal{M}$ and $\widetilde{PT^*\mathcal{M}}$ as the unit map of $\mathbf{PT}^*\mathcal{M} \cat \mathcal{M}$, 
it becomes a smooth precategory.
Bisections of $\mathbf{PT}^*\mathcal{M}$ are defined similarly through the ones in
the precategories $PT^*\mathcal{M}, \widetilde{PT^*\mathcal{M}}$, namely, 
they are diffeomorphisms associated with paths in $\mathcal{M}$ and $T\mathcal{M}$.
Explicitly, for bisections $\Sigma = (\phi,\eta)$, $\tilde{\Sigma} =
(\tilde{\phi}, \tilde{\eta})$ of $PT^*\mathcal{M}$ and $\widetilde{PT^*\mathcal{M}}$, a bisection
$\mathbf{\Sigma}$ of $\mathbf{PT}^*\mathcal{M}$ is given by $\mathbf{\Sigma} = \Sigma \diamond \tilde{\Sigma}$.

We then define a pre-rack product $\prack$ in the precategory
$\mathbf{PT}^*\mathcal{M} \cat \mathcal{M}$.
We propose a product of bisections between
$\mathbf{\Sigma} = \tilde{\Sigma} \diamond \Sigma$ and $\mathbf{T} = \tilde{T} \diamond T$ 
of $\mathbf{PT}^*\mathcal{M} \cat \mathcal{M}$ as 
\begin{align}
\mathbf{\Sigma} \prack \mathbf{T}
=& \
\Bigl(
\phi_1 \circ \psi_t \circ \phi_1^{-1}
\diamond
\tilde{\phi}_1 \circ \tilde{\psi}_t \circ \tilde{\phi}_1^{-1}
\,
,
\notag \\
& \qquad 
(
(\phi_1^{-1})^* (\zeta_t) - (\phi_1^{-1})^* \iota_{\dot{\psi}} \phi_1^*
 \mathrm{d} \beta_{\Sigma}
)
+
(
(\tilde{\phi}_1^{-1})^* (\tilde{\zeta}_t)
-
(\tilde{\phi}^{-1})^*_1 \tilde{\iota}_{\dot{\tilde{\psi}}}
 \tilde{\phi}_1^* \tilde{\mathrm{d}}
 \tilde{\beta}_{\tilde{\Sigma}}
)
\Bigr).
\label{eq:pre-rack_product}
\end{align}
Here 
\begin{align}
\beta_{\Sigma} = \int^1_0 \! ds \ \phi_s^* \eta_s, \qquad 
\tilde{\beta}_{\tilde{\Sigma}} = \int^1_0 \! ds \ \tilde{\phi}_s \tilde{\eta}_s
\end{align}
are 1-forms associated with the bisections $\Sigma, \tilde{\Sigma}$.
We note that the product defined by \eqref{eq:pre-rack_product} is quite
different from the rack product in
\eqref{eq:cotangent_path_rack_product}.
Remarkably, in the cotangent path in
\eqref{eq:cotangent_path_rack_product}, the pull-back $\phi^*$ induced
by the path $\phi$ in the base space acts on the 1-form by the rack product.
In \eqref{eq:pre-rack_product}, it is not true that the pull-back
$(\tilde{\phi} \diamond \phi)^{-1 *}$ by the path $\tilde{\phi} \diamond
\phi$ in $\mathcal{M}$ acts on the doubled 1-form, given 
in the form of $X^{\mu} d x^{\mu} + \xi_{\mu} d \tilde{x}^{\mu}$, like that way.
More explicitly, the pre-rack product of $\mathbf{\Sigma}$ and
$\mathbf{T}$ at $(x,\tilde{x}) \in
\mathcal{M}$ is given by 
\begin{align}
(\mathbf{\Sigma} \prack \mathbf{T})_t (x,\tilde{x})
=& \ 
\Bigl(
(
\phi_1 \circ \psi_t \circ \phi_1^{-1} (x), \,
\tilde{\phi}_1 \circ \tilde{\psi}_t \circ \tilde{\phi}_1^{-1}
 (\tilde{x})
)
, 
\notag \\
& \qquad  
(
(\phi_1^{-1})^* (\zeta_t) - (\phi_1^{-1})^* \iota_{\dot{\psi}} \phi_1^*
 \mathrm{d} \beta_{\Sigma} 
)
(x, \tilde{x})
+
(
(\tilde{\phi}_1^{-1})^* (\tilde{\zeta}_t)
-
(\tilde{\phi}^{-1})^*_1 \tilde{\iota}_{\dot{\tilde{\psi}}}
 \tilde{\phi}_1^* \tilde{\mathrm{d}}
 \tilde{\beta}_{\tilde{\Sigma}} 
)
(x, \tilde{x})
\Bigr).
\end{align}
One notices that there are no terms such as $(\phi_1^{-1})^* \tilde{\zeta}_t$.
Here the first line provides the components of path in $\mathcal{M}$.
The path is represented by a pair of local coordinates $(x,\tilde{x})$.
The second line gives the components of the doubled cotangent vectors in
the form of $X^{\mu} d x^{\mu} + \xi_{\mu} d \tilde{x}^{\mu}$ on $T^* \mathcal{M}$.
The maps $\phi, \psi$ controls the translation along $x$-direction while
$\tilde{\phi}, \tilde{\psi}$ gives the path in the $\tilde{x}$-direction.
With these structures, we have the following proposition.
\begin{proposition}[Pre-rackoid by doubled cotangent path]
The pre-rack product $\prack$ defined in \eqref{eq:pre-rack_product}
 does not satisfy the self-distributivity in general.
Therefore $(\mathbf{PT}^*\mathcal{M} \cat \mathcal{M}, \prack)$ is a pre-rackoid.
We call this the doubled cotangent path pre-rackoid.
\end{proposition}
The non-self-distributivity of the pre-rack product
\eqref{eq:pre-rack_product} is shown by direct
calculations.
However we present a concise reason in the following.
The path in the base space is nothing but the one in the path rackoid.
Since the rack action along the $x$($\tilde{x}$)-direction is given by the
adjoint action
\begin{align}
(\phi \rack \psi)_t = \phi_1 \circ \psi_t \circ \phi_1^{-1},
\end{align}
then, it is obvious that this satisfies the self-distributivity.
The cotangent part seems to be less trivial.
For example, the 1-form $\beta_{\mathbf{\Sigma} \prack \mathbf{T}}$
associated with $\mathbf{\Sigma} \prack \mathbf{T}$ is given by 
\begin{align}
\beta_{\mathbf{\Sigma} \prack \mathbf{T}} 
=& \ 
\int^1_0 \! ds 
\left[
\phi_1 \circ \psi_s \circ \phi_1^{-1} \diamond \tilde{\phi}_1 \circ
 \tilde{\psi}_t \circ \tilde{\phi}_1^{-1}
\right]^* 
\notag \\
& \times
\left[
(
(\phi_1^{-1})^* (\zeta_t) - (\phi_1^{-1})^* \iota_{\dot{\psi}} \phi_1^*
 \mathrm{d} \beta_{\Sigma}
)
+
(
(\tilde{\phi}_1^{-1})^* (\tilde{\zeta}_t)
-
(\tilde{\phi}^{-1})^*_1 \tilde{\iota}_{\dot{\tilde{\psi}}}
 \tilde{\phi}_1^* \tilde{\mathrm{d}}
 \tilde{\beta}_{\tilde{\Sigma}}
)
\right].
\end{align}
Here one finds that 
$
\left[
\phi_1 \circ \psi_s \circ \phi_1^{-1} \diamond \tilde{\phi}_1 \circ
 \tilde{\psi}_t \circ \tilde{\phi}_1^{-1}
\right]^*
$,
composed of the pull-back of $\phi_1$ along $x$-direction, cancels $(\phi_1^{-1})^*$
in front of $\zeta_t$, but that for the $\tilde{x}$-direction pics up
shifts of the path by $\tilde{\phi}_1^*$.
Moreover, the $\mathrm{d}$ operation not only acts on the cotangent path in $\mathcal{F}_{x,[\tilde{x}]}$
but also on that in $\tilde{\mathcal{F}}_{[x],\tilde{x}}$. 
The same is true for $\mathrm{d}^*$.
Due to these properties, the expression \eqref{eq:d_1-form} does not hold anymore
and the proof in \cite{Laurent-Gengoux:2018zoz} is not applied to the product \eqref{eq:pre-rack_product}.
These properties trigger the breaking of the self-distributivity.

The remaining discussion is completely parallel to the ones in
\cite{Laurent-Gengoux:2018zoz}.
Since the pre-rack action is smooth, the adjoint map induced by the $\prack$
results in a bracket on the infinitesimal algebroid $\mathcal{A}$ of
$(\mathbf{PT}^*\mathcal{M} \cat M, \prack)$.
Again, by differentiating the families of bisections in the base space, we obtain a pair of vectors:
\begin{align}
\frac{\partial}{\partial u} \phi^u |_{u=0} = X = X^{\mu} \partial_{\mu},
 \quad
\frac{\partial}{\partial u} \tilde{\phi}^u |_{u=0} = \xi = \xi_{\mu} \tilde{\partial}_{\mu}.
\end{align}
Similarly, for families of bisections in the cotangent bundle, we obtain a pair of 1-forms:
\begin{align}
\frac{\partial}{\partial u} \beta_{\Sigma^u} |_{u=0} = \alpha = \alpha_{\mu}
 dx^{\mu},
\quad 
\frac{\partial}{\partial u} \beta_{\tilde{\Sigma}^u} |_{u=0} = A =
 A^{\mu} d \tilde{x}_{\mu}.
\end{align}
We here clarify the relation of the doubled geometry and generalized geometry.
The vectors $\xi_{\mu} (x,\tilde{x}) \tilde{\partial}^{\mu}$ and
1-forms $X^{\mu} (x,\tilde{x}) d \tilde{x}_{\mu}$ on $T\mathcal{M}_-, (T\mathcal{M}_-)^*$
in the doubled geometry are identified with the 1-forms $\xi_{\mu}
(x,\tilde{x}) dx^{\mu}$ and vectors $X^{\mu} (x,\tilde{x})
\partial_{\mu}$ on $T\mathcal{M}_+, (T\mathcal{M}_+)^*$ through the following natural isomorphism \cite{Freidel:2017yuv}:
\begin{align}
\Phi^+ \ : \ \xi_{\mu} \tilde{\partial}^{\mu} \sim \xi_{\mu} d x^{\mu},
 \quad 
X^{\mu} d \tilde{x}_{\mu} \sim X^{\mu} \partial_{\mu}.
\label{eq:natural_isomorphism}
\end{align}
Then, the pair of paths $(\phi, \tilde{\eta})$ on $\mathcal{M}$ and pair
of vectors on $T\mathcal{M}_+$ defined by the derivatives of the paths,
and the pair of 1-forms $(\eta, \tilde{\phi})$ on the dual bundle $(T\mathcal{M}_+)^*$
are obtained. 
Given these identifications, 
it is now straightforward to obtain the D-bracket structure from the the pre-rack action \eqref{eq:pre-rack_product}.
One finds that the Dorfman bracket of the standard Courant algebroid
comes from the $PT^*\mathcal{M} \cat \mathcal{F}_{x,[\tilde{x}]}$ part in \eqref{eq:pre-rack_product}.
The extra terms needed for the D-bracket of the Vaisman algebroid is
obtained from the $\widetilde{PT^*\mathcal{M}} \cat \tilde{\mathcal{F}}_{[\phi(x)],\tilde{x}}$ part.
As we have seen in the case of the standard Courant algebroid, the map
$\varphi = (\mathrm{ev}_1,\int_0^1 \! dt)$ finally provides the complete
D-bracket 
\begin{align}
\vbracket{e_1}{e_2}_D =& \ 
[X_1, X_2]_{T\mathcal{M}_+}
+ \mathcal{L}_{\xi_1} X_2 - \iota_{\xi_2} \mathrm{d}^* X_1
\notag \\
+ & \ 
[\xi_1, \xi_2]_{(T\mathcal{M}_+)^*}
+ \mathcal{L}_{X_1} \xi_2 - \iota_{X_2} \mathrm{d} \xi_1.
\label{eq:D-bracket2}
\end{align}
By definition, the infinitesimal algebroid $\mathcal{A}$ of the
pre-rackoid $\mathbf{PT}^*\mathcal{M} \cat \mathcal{M}$ equipped with the bracket
\eqref{eq:D-bracket2} is the Vaisman algebroid.

When the strong constraint is imposed, the Leibniz identity of the
Vaisman algebroid is recovered and it becomes the Courant algebroid.
This is obvious if one employs a solution as the one that
$\tilde{\partial}^{\mu} * = 0$, namely, the para-holomorphic quantities.
In this case, the physical spacetime is defined by a leaf $\mathcal{F}_{x,[\tilde{x}]}$ for
a fixed $\tilde{x}$ and all the physical quantities depend only on $x$.
The directions along the leaf $\tilde{\mathcal{F}}_{[x],\tilde{x}}$ is frozen in the
doubled space.
Then, the $\widetilde{PT^*\mathcal{M}}$ component of the pre-rackoid $\mathbf{PT}^*\mathcal{M} \cat
\mathcal{M}$ becomes trivial and it reduces to the rackoid $PT^*\mathcal{M}
\cat \mathcal{F}_{x,[\tilde{x}]} \subset \mathcal{M}$.
The tangent space of $\mathcal{M}$ is identified with the generalized
tangent space $\mathbb{T}M$ through the natural isomorphism \eqref{eq:natural_isomorphism}.
At the level of the algebroid, the condition $\tilde{\partial}^{\mu} * =
0$ implies $[\cdot, \cdot]_{*} = \mathcal{L}_{\xi} * = \mathrm{d}^* = 0$
and the D-bracket reduces to the Dorfman bracket \cite{Mori:2019slw}.

Now we revisit the geometrical meaning of the strong constraint in DFT.
As we have shown above, an infinitesimal algebroid of the pre-rackoid is
isomorphic to the Vaisman algebroid equipped with the D-bracket, which
is equivalent to the C-bracket through the anti-symmetrization.
An essential difference between the Vaisman and the Courant algebroids
is the Leibniz (Jacobi) identity of the D-bracket (C-bracket).
As we have mentioned, the Leibniz identity is nothing but the
self-distributivity of the rack action.
From the viewpoint of the doubled geometry, the strong constraint can be
seen as a condition for the recovery of the self-distributivity of the pre-rackoid.

This fact is rephrased in the following suggestive form.
We define the operator 
$R(g \otimes h) = g \otimes g \rack h$ 
on $\text{Bis}(\mathbf{PT}^*\mathcal{M}) \otimes \text{Bis}(\mathbf{PT}^*\mathcal{M})$ for any $g,h \in \text{Bis}(\mathbf{Y})$.
Then the action of $R$ on the tensor products $\text{Bis}(\mathbf{PT}^*\mathcal{M}) \otimes \text{Bis}(\mathbf{PT}^*\mathcal{M}) \otimes
\text{Bis}(\mathbf{PT}^*\mathcal{M})$ results in 
\begin{align}
&
R_{12} R_{13} R_{23} 
(g,h,i)
=
g \otimes g \rack h \otimes g \rack (h \rack i),
\notag \\
&
R_{23} R_{13} R_{12}
(g,h,i)
=
g \otimes g \rack h \otimes (g \rack h) \rack (g \rack i).
\end{align}
Therefore the self-distributivity of the doubled cotangent path in $\mathcal{M}$
is recast in the following quantum Yang-Baxter equation:
\begin{align}
R_{12} R_{13} R_{23} = R_{23} R_{13} R_{12}.
\end{align}
In other words, we can say that the strong constraint in DFT is an sufficient
condition of the quantum Yang-Baxter equation for the rack action.

\section{Formal rackoids and pre-rackoids} \label{sct:formalrackoid}
In the previous section, we work on the integration of the Vaisman algebroid by a heuristic approach based on the doubled geometry.
In this section, we propose a formal (pre)-rackoids which enable one to find a formal integration of the Courant and the Vaisman algebroid.
The prescription discussed here is useful for perturbative treatment of the (pre)-rackoids.

\subsection{Formal rackoids}
Let $\mathfrak{g}$ be a (left) Leibniz algebra with a Leibniz bracket 
$[-,-]$. Let $\mathrm{ad}(X)Y = [X, Y]$.
We define an operation \cite{Kinyon:2004},
\begin{eqnarray}
X \rhd Y := \exp \mathrm{ad}(X)Y.
\end{eqnarray}
Then, the operation satisfies
\begin{eqnarray}
X \rhd (Y \rhd Z) &=& (X \rhd Y) \rhd (X \rhd Z),
\end{eqnarray}
since 
\begin{eqnarray}
X \rhd (Y \rhd Z) &=& \exp \mathrm{ad}(X) (\exp \mathrm{ad}(Y) Z)
\nonumber \\
&=& \exp (\exp \mathrm{ad}(X) \mathrm{ad}(Y)) \exp \mathrm{ad}(X) Z
\nonumber \\
&=& \exp (\mathrm{ad}( \exp \mathrm{ad}(X) Y)) \exp \mathrm{ad}(X) Z
\nonumber \\
&=& (X \rhd Y) \rhd (X \rhd Z).
\end{eqnarray}

We generalize this construction to a Leibniz algebroid, Courant algebroid and a Vaisman algebroid.

We introduce a formal integration of a Leibniz algebroid.
\begin{define}
Let $E$ be a vector bundle over $M$. For $e_i \in \Gamma(E)$ and $f \in C^{\infty}(M)$, we consider a product $\cdot: \Gamma(E) \times C^{\infty}(M) \rightarrow C^{\infty}(M)$, and operations, a rack operation,
$\rhd: \Gamma(E) \times \Gamma(E) \rightarrow \Gamma(E)$ and 
a rack action, $\rhd: \Gamma(E) \times C^{\infty}(M) \rightarrow C^{\infty}(M)$.
If they satisfy
\begin{eqnarray}
e_1 \rhd (e_2 \rhd e_3) &=& (e_1 \rhd e_2) \rhd (e_1 \rhd e_3) ,
\label{rackidentity1}
\\
e_1 \rhd (e_2 \rhd f) &=& (e_1 \rhd e_2) \rhd (e_1 \rhd f),
\label{rackidentity2}
\\
e_1 \rhd fe_2 &=& (e_1 \rhd f) \cdot (e_1 \rhd e_2),
\label{rackidentity3}
\end{eqnarray}
$(E, \rhd, \cdot)$ is called a \textit{bundle (Lie) rackoid}.
\end{define}
In fact a bundle rackoid is constructed by the formal exponential of operations of the Leibniz algebroid.
Let $(E, [-,-]_D, \rho)$ be a Leibniz algebroid over a smooth manifold $M$.
Then, for sections $e_i \in \Gamma(E)$ and a function $f \in C^{\infty}(M)$, 
we define a rack operation $\rhd: \Gamma(E) \times \Gamma(E) \rightarrow \Gamma(E)$ and a rack action on $C^{\infty}(M)$ 
$\rhd: \Gamma(E) \times C^{\infty}(M) \rightarrow C^{\infty}(M)$ as
\begin{eqnarray}
e_1 \rhd e_2 &:=& \exp \mathrm{ad}(e_1)e_2,
\label{formal1}
\\
e_1 \rhd f &:=& (\exp \rho(e_1)) f.
\label{formal2}
\end{eqnarray}
Here $\mathrm{ad}(e_1)e_2 = [e_1, e_2]_D$. 
Then, we obtain the rack identities,
\eqref{rackidentity1}--\eqref{rackidentity3} 
from the following three identities of a Leibniz algebroid,
\begin{eqnarray}
[e_1, [e_2, e_3]_D]_D &=& [[e_1, e_2]_D, e_3]_D + [e_2, [e_1, e_3]_D]_D,
\nonumber \\
\rho([e_1, e_2]_D)f &=& [\rho(e_1), \rho(e_2)] f,
\nonumber \\
~[e_1, f e_2]_D &=& f [e_1, e_2] + \rho(e_1) f \cdot e_2.
\nonumber
\end{eqnarray}
We call a \textit{formal rackoid} a rackoid defined by 
formal exponentials \eqref{formal1} and \eqref{formal2}
of a tangent algebroid.

\begin{theorem}
If $E$ is a Leibniz algebroid with $(\rho, [-,-]_D)$, 
$E$ is a formal rackoid,
if we define
\begin{eqnarray}
e_1 \rhd e_2 &:=& \exp \mathrm{ad}(e_1)e_2,
\\
e_1 \rhd f &:=& (\exp \rho(e_1)) f.
\end{eqnarray}
\end{theorem}
Next, we consider a bundle rackoid corresponding to a Courant algebroid.
\begin{define}
Let $(E, \rhd, \cdot)$ be a bundle rackoid.
Let $(\cdot,\cdot)$ be a pseudo-Euclidean inner product on $E$.
If a bundle rackoid satisfies
\begin{eqnarray}
e_1 \rhd (e_2, e_3) &=& ((e_1 \rhd e_2), (e_1 \rhd e_3)),
\label{rackidentity4}
\end{eqnarray}
$E$ is called a \textit{metric bundle rackoid}.
\end{define}
Let $(E, [-,-]_D, \rho, (\cdot,\cdot))$ be a Courant algebroid. We consider a formal rackoid defined by operations \eqref{formal1} and \eqref{formal2}. Then, the condition \eqref{rackidentity4} is proved from the identity of the Courant algebroid,
\begin{eqnarray}
\rho(e_1)(e_2, e_3) &=& ([e_1, e_2]_D, e_3) + (e_2, [e_1, e_3]_D).
\end{eqnarray}
This identity with identities as Leibniz algebroid is enough to obtain other identities of the Courant algebroid.
Then, we obtain the following theorem.
\begin{theorem}
If $E$ is a Courant algebroid with $([-,-]_D, \rho, (\cdot,\cdot))$, 
If we define a formal rackoid with operations,
\begin{eqnarray}
e_1 \rhd e_2 &:=& \exp \mathrm{ad}(e_1)e_2,
\\
e_1 \rhd f &:=& (\exp \rho(e_1)) f,
\end{eqnarray}
$E$ is a metric bundle rackoid.
\end{theorem}

\subsection{Formal pre-rackoids}
We consider a pre-rackoid version of a formal exponential.

\begin{define}
Let $E$ be a vector bundle over a smooth manifold $M$ with a pseudo-Euclidean inner product $(\cdot,\cdot)$.
For $e_i \in \Gamma(E)$ and $f \in C^{\infty}(M)$, we consider a product $\cdot: \Gamma(E) \times C^{\infty}(M) \rightarrow C^{\infty}(M)$, and operations, a rack operation,
$\prack: \Gamma(E) \times \Gamma(E) \rightarrow \Gamma(E)$ and 
a rack action, $\prack: \Gamma(E) \times C^{\infty}(M) \rightarrow C^{\infty}(M)$.

If an operation $\prack$ satisfies
\begin{eqnarray}
e_1 \prack fe_2 &=& (e_1 \prack f)\cdot(e_1 \prack e_2),
\label{rackidentity13}
\\
e_1 \prack (e_2, e_3) &=& ((e_1 \prack e_2), (e_1 \prack e_3)),
\label{rackidentity14}
\end{eqnarray}
$E$ is called a \textit{metric (bundle) (Lie) pre-rackoid}.
\end{define}

Let $(E, \vbracket{-}{-}_D, \rho, (\cdot, \cdot))$ be a Vaisman algebroid.
As in the case of a Leibniz algebroid, 
we define the following operation $\prack: \Gamma(E) \times \Gamma(E) \rightarrow \Gamma(E)$ and the action on $C^{\infty}(M)$ 
$\prack: \Gamma(E) \times C^{\infty}(M) \rightarrow C^{\infty}(M)$
for $e_i \in \Gamma(E)$ and $f \in C^{\infty}(M)$, 
\begin{eqnarray}
e_1 \prack e_2 &:=& \exp \mathrm{ad}(e_1)e_2,
\label{formal3}
\\
e_1 \prack f &:=& \exp \rho(e_1) f.
\label{formal4}
\end{eqnarray}
Here $\mathrm{ad}(e_1)e_2 = \vbracket{e_1}{e_2}_D$ is a D-bracket.
The exponential of a D-bracket or a generalized Lie derivative 
$\exp \calL_{v}$ has appeared as a large gauge transformation of DFT 
\cite{Hohm:2012gk}.
We can prove that operations \eqref{formal3} and 
\eqref{formal4} satisfy \eqref{rackidentity13} and \eqref{rackidentity14}.
A Vaisman algebroid gives a formal metric pre-rackoid.
In general, the rack identities are not satisfied,
\begin{eqnarray}
e_1 \prack (e_2 \prack e_3) & \neq & (e_1 \prack e_2) \prack (e_1 \prack e_3) ,
\label{rackidentity21}
\\
e_1 \prack (e_2 \prack f) & \neq & (e_1 \prack e_2) \prack (e_1 \prack f),
\label{rackidentity22}
\end{eqnarray}
The closure condition is expressed by
\begin{eqnarray}
&& e_1 \prack (e_2 \prack e_3) - (e_1 \prack e_2) \prack (e_1 \prack e_3) = 0,
\label{rackidentity31}
\\
&& e_1 \prack (e_2 \prack f) - (e_1 \prack e_2) \prack (e_1 \prack f) = 0,
\label{rackidentity32}
\end{eqnarray}

\section{Rackoids and pre-rackoids from sigma models} \label{sct:sigma_models}
In this section, we consider a sigma model description of a cotangent path rackoid and a doubled cotangent path pre-rackoid.
We discuss description of a Lie rackoid and pre-rackoid using topological 
sigma models.

\subsection{Courant sigma model}\label{CSM}
We can naturally construct a three dimensional topological sigma model with a 
Courant algebroid structure called a Courant sigma model.
\cite{Ikeda:2002wh, Roytenberg:2006qz}

Let $(E, \rho, [-,-]_D, H)$ be a Courant algebroid over a smooth manifold $M$.
Let $N$ be a three dimensional manifold with local coordinates $\sigma^{\mu}$.
$\varphi: N \rightarrow M$ is a smooth map, $A \in \Omega^1(N, \varphi^*E)$ is a 1-form and $B \in \Omega^2(N, \varphi^*T^*M)$ is a 2-form. 

The action of the Courant sigma model is
\begin{align}
S &= \int_{N} \left(- B_i \wedge d \varphi^i + \frac{1}{2} k_{ab} A^a \wedge dA^b
+ \rho^i{}_a(\varphi) B_i \wedge A^a + \frac{1}{3!} H_{abc}(\varphi) A^a \wedge A^b \wedge A^c\right),
\label{actionCSM}
\end{align}
where $k_{ab}$ is defined from a fiber metric as $k_{ab} = \bracket{e_a}{e_b}$,
$\rho(e_a) = \rho^i{}_a(x)\partial_i = \rho^i{}_a(x)\frac{\partial}{\partial x^i}$ and $\frac{1}{3!} H_{abc}(x) = H(e_a, e_b, e_c)$. Here $e_a$ is a basis of the fiber of $E$.

The gauge transformation is
\begin{eqnarray}
\delta \varphi^i &=& \rho^i{}_a(\varphi) t^a,
\label{gfofCSM1}
\\
\delta A^a &=& d t^a + k^{ab} \rho^i{}_b(\varphi) u_i + k^{ab} H_{bcd}(\varphi) A^c t^d,
\label{gfofCSM2}
\\
\delta B_i &=& d u_i + \partial_i \rho^j_a (A^a u_j - t^a B_j)
+ \frac{1}{2} \partial_i H_{abc}(\varphi) A^a A^b t^c,
\label{gfofCSM3}
\end{eqnarray}
where $t^a$ is a 0-form gauge parameter and $u_i$ is a 1-form gauge parameter.
The action \eqref{actionCSM} is gauge invariant under the gauge transformation
\eqref{gfofCSM1}--\eqref{gfofCSM3} if and only if the target space $E$ with structures $(k, \rho, H)$ is a Courant algebroid. Thus this topological sigma mode is called a Courant sigma model.

We consider the case of the standard Courant algebroid on $E= TM \oplus
T^*M$. We take the $O(D,D)$ metric as
\begin{eqnarray}
  k = \left(
    \begin{array}{cc}
      0 & 1  \\
      1 & 0 
    \end{array}
  \right)
\label{fibermetric}
\end{eqnarray}
and the anchor map $\rho: X + \alpha \mapsto X$, where $X + \alpha \in \Gamma(TM\oplus T^*M)$, 
\begin{eqnarray}
  \rho = \left(
    \begin{array}{cc}
      1 & 0 \\
      0 & 0 
    \end{array}
  \right)
\label{standardanchor}
\end{eqnarray}
and $H =0$.
The 1-form field $A^a$ is decomposed to components of $TM$ and $T^*M$ as $A^a = (A^i, C_i)$,
where $A \in \Omega^1(N, \varphi^*T^*M)$ and $C \in \Omega^1(N, \varphi^*TM)$.
The action of the Courant sigma model \eqref{actionCSM} becomes a simple form,
\begin{eqnarray}
S &=& \int_{N} - B_i \wedge d \varphi^i + C_i \wedge dA^i + B_i \wedge A^i.
\end{eqnarray}

Let $N = \Sigma \times \bR$, where $\Sigma$ is a two dimensional manifold and $\bR$ is the time direction.
Then, we can compute the symplectic form,
\begin{eqnarray}
\omega = \int_{\Sigma} \delta X^i \wedge \delta B_i + \delta A^i \wedge \delta C_i.
\end{eqnarray}
Nonzero Poisson brackets of fields are obtained from Poisson brackets of canonical conjugates,
\begin{eqnarray}
\{X^i(\sigma), B_{abj}(\sigma^{\prime}) \} &=& - \epsilon_{ab} \delta^i_j
\delta^2 (\sigma - \sigma^{\prime}),
\label{Poissonbracket1}
\\
\{A^i_a(\sigma), C_{bj}(\sigma^{\prime}) \} &=& \epsilon_{ab} \delta^i_j \delta^2 (\sigma - \sigma^{\prime}),
\label{Poissonbracket2}
\end{eqnarray}
where $a, b = 1, 2$ are indices on $\Sigma$.
The Hamiltonian 
\begin{eqnarray}
H &=& \int_{\Sigma} (B_{0i} \wedge G^i - C_{0i} \wedge F^i - A_0^i K_{21i}), 
\end{eqnarray}
is purely written by terms with constraints,
where $B_{0i} = B_{0ai} d\sigma^a$ where $a=1,2$.
Here constraints are
\begin{eqnarray}
G^i &=& d \varphi^i - A^i,
\\
F_{i} &=& d C_i + B_i,
\\
K_{i} &=& d A^i.
\end{eqnarray}
Poisson brackets of constraints $G^i$, $F_{i}$ and $K_{i}$ show that 
they consist of the first class constraints.

\subsection{Path rackoids from Courant sigma models}\label{PathrackoidCSM}
We analyze correspondence of the standard Courant sigma model and a path Lie rackoid. 

We consider a path $I_t = [0, t] \subset \Sigma$ and a map from $I_t$ to the target space $M$, $\gamma(t):I_t \rightarrow M$, where $t \in \bR$.
$a=(\gamma, \alpha) \in PT^*M$ consists of a cotangent path on $M$.

We can easily identify a path to a map $\varphi$ on $I_t$, $\gamma = \varphi|_{I_t}$.
A section of a generalized tangent bundle $TM \oplus T^*M$,
$X  + \alpha = X^i(x) \partial_i + \alpha_i(x) dx^i$,
is mapped as follows,
\begin{eqnarray}
(\tilx + \tila) = \int_0^1 (X^i(\varphi(\sigma)) C_i(\sigma) + \alpha_i(\varphi(\sigma)) A^i(\sigma)),
\end{eqnarray}
using fields  $A^i(\sigma)$ and $C_i(\sigma)$ in the Courant sigma model.
This map is denoted by $j$,
\begin{eqnarray}
&& j: X + \alpha \mapsto \tilx + \tila,
\end{eqnarray}
Then the bilinear form $(\cdot, \cdot)$ is mapped to a Poisson bracket since
$j$ is a homomorphism,
\begin{eqnarray}
&& j:(X + \alpha, Y + \beta) \mapsto - \{ \tilx + \tila, \tily + \tilb \}.
\label{operation1}
\end{eqnarray}
for $X+ \alpha, Y + \beta \in \Gamma(TM \oplus T^*M)$.
The anchor map $\rho(X + \alpha) = X$ is mapped to 
\begin{eqnarray}
&& j:\rho(X + \alpha) f(x) \mapsto - \{ \{ \tilx + \tila, H \}, f(\varphi(\sigma) \}.
\label{operation2}
\end{eqnarray}
where $f \in C^{\infty}(M)$.
The Dorfman bracket is mapped to the derived bracket of Poisson brackets,
\begin{eqnarray}
&& j:[X + \alpha, Y+\beta]_D \mapsto - \{ \{ \tilx + \tila, H \}, \tily + \tilb \}.
\label{operation3}
\end{eqnarray}
where $f \in C^{\infty}(M)$. 
Such a derived bracket construction has been formulated in 
\cite{Roytenberg:1999}.
We can prove that $j$ is a homomorphism of a Courant algebroid.
Therefore a Courant algebroid structure on $E$ is mapped to the mapping space
$\Map(N, E)$ with a symplectic form $\omega$, and operations are calculated by \eqref{operation1}--\eqref{operation3}.

From the above correspondence, a cotangent path is mapped to
\begin{eqnarray}
&& j: (\gamma, \alpha) \mapsto (\varphi(t), \tila(t)).
\end{eqnarray}
Quantities $\phi, \psi, \beta, \zeta$ in a rackoid which satisfy 
differential equations \eqref{differentialequations} are 
formally described by exponential maps, which are 
Wilson lines
\begin{eqnarray}
j:\phi &\mapsto& \mathrm{P} \exp \tilx = \mathrm{P} \exp \int_{\gamma} X^i(\varphi(\sigma)) C_i(\sigma), 
\label{wilson1}
\\
j:\beta &\mapsto& \mathrm{P} \exp \tila = \mathrm{P} \exp \int_{\gamma} \alpha_i(\varphi(\sigma)) A^i(\sigma)),
\label{wilson2}
\end{eqnarray}
Here $\gamma = \gamma(1)$. 
We obtain a formal rackoid structure in section \ref{sct:formalrackoid} from  Wilson lines \eqref{wilson2} and \eqref{wilson2}. Wilson lines provide formal exponential maps, but they are useful for concrete calculations of the integration using the quantization of the sigma model.

We can easily generalize the above construction of the standard Courant algebroid to a general Courant algebroid. The equivalent but more familiar construction for quantization of a Courant algebroid using a topological sigma model is a so called AKSZ sigma model using supergeometry. 
\cite{AKSZ, Ikeda:2012pv}

\subsection{Topological double sigma models and pre-rackoids}
In this section, we consider a construction of a pre-rackoid using
a topological sigma model of doubled target spacetime
with a Vaisman algebroid structure.
A topological sigma model with a structure of DFT geometry is proposed in
\cite{Kokenyesi:2018xgj}.

Let $\wM$ is $2D$-dimensional manifold corresponding a doubled spacetime.
Typically, it is a direct product of a physical spacetime and a T-dual spacetime, $\wM = M \times \widetilde{M}$.
Suppose an $O(D,D)$ invariant metric $\eta_{IJ}$ on $\wM$, where $I, J = 1, \cdots 2D$ are indices of local coordinate on $\wM$.

Let $N$ be a 3 dimensional manifold with local coordinates $\sigma^{\mu}$.
$\varphi: N \rightarrow \wM$ is a smooth map from $N$ to the target double spacetime.
$A \in \Omega^1(N, \varphi^*T^*\wM)$ is a 1-form taking a value on the pullback of 
$T^*\wM$. $B \in \Omega^2(N, \varphi^*T^*M)$ is a 2-form taking a value on the pullback of $T^*\wM$.

An action of a topological double sigma model is 
\begin{eqnarray}
S &=& \int_{N} \left(- B_I d \varphi^I + \frac{1}{2} \eta^{IJ} A_I d A_J + \eta^{IJ} B_I A_J + \frac{1}{3!} F^{IJK}(\varphi) A_I A_J A_K\right),
\label{actionTDSM}
\end{eqnarray}
where $F(x) \in \Omega^3 (\wM)$ is a generalized flux which is a 3-form on $\wM$. 
Now we consider a simplest $F=0$ case for a genuine Vaisman algebroid. Then the action becomes
\begin{eqnarray}
S &=& \int_{N} (- B_I d \varphi^I + \frac{1}{2} \eta^{IJ} A_I d A_J + \eta^{IJ} B_I A_J).
\end{eqnarray}
Nonzero Poisson brackets of fields are computed as
\begin{eqnarray}
\{X^I(\sigma), B_{abJ}(\sigma^{\prime}) \} &=& - \epsilon_{ab} \delta^I_J
\delta^2 (\sigma - \sigma^{\prime}),
\label{Poissonbracket11}
\\
\{A_{aI}(\sigma), A_{bJ}(\sigma^{\prime}) \} &=& \epsilon_{ab} \eta_{IJ} \delta^2 (\sigma - \sigma^{\prime}),
\label{Poissonbracket12}
\end{eqnarray}
where $a, b = 1, 2$ are indices on $\Sigma$.

The section condition to reduce the 2D-dimensional double spacetime $\wM$ to a D-dimensional spacetime $M$ drop functions of $\phi^I$. If we use equations of motion, half degrees of $A_I$ drop and we obtain the first class constraint $G^I|_M$. On the other hand, we can impose $G^I$ consist of the first class constraints. The condition gives the section condition on fields on the target space.

It is natural to consider the following gauge transformations 
analogous to gauge transformations of the Couorant sigma model, \eqref{gfofCSM1}--\eqref{gfofCSM3},
\begin{eqnarray}
\delta \varphi^I &=& \eta^{IJ} t_J,
\\
\delta A_I &=& d t_I + u_I, 
\\
\delta B_I &=& d u_I,
\end{eqnarray}
where $t_I$ is a zero-form gauge parameter 
and $u_I$ is a 1-form gauge parameter. 
However, the action $S$ is not gauge invariant under this transformation,
$\delta S \neq 0$.

In the Hamiltonian analysis, we obtain a Hamiltonian,
\begin{eqnarray}
H &=& \int_{\Sigma} (B_{0I} \wedge G^I + A_{0}^I F_{I}), 
\end{eqnarray}
where constraints are
\begin{eqnarray}
G^I &=& d \varphi^I - \eta^{IJ} A_{J},
\\
F_{I} &=& d A_{I} + B_{I}.
\end{eqnarray}
In this case, the constraints are not the first class, since
\begin{eqnarray}
\{ G_a^I(\sigma), G_b^J(\sigma) \} &=& \eta^{IJ} \epsilon_{ab} \delta^2 (\sigma - \sigma).
\end{eqnarray}
where $a, b = 1, 2$ are indices on $\Sigma$.

Operations in the Vaisman algebroid are constructed similar to  one of the Courant algebroid discussed in sections \ref{CSM} and \ref{PathrackoidCSM}.

A section of tangent bundle $T\wM$,
$X= X^I(x) \partial_I$ is mapped to a field (a pullback 1-form to the mapping space),
\begin{eqnarray}
\tilx(t) = \int_{I_t} X^I(\varphi(\sigma)) A_I(\sigma).
\end{eqnarray}
The map is denoted by $j$,
\begin{eqnarray}
&& j: X \mapsto \tilx,
\end{eqnarray}
The bilinear form $(\cdot, \cdot)$ is mapped to a Poisson bracket since
$j$ is a homomorphism,
\begin{eqnarray}
&& j:(X, Y) \mapsto - \{ \tilx, \tily \}.
\label{operation11}
\end{eqnarray}
for $X, Y \in \Gamma(T\mathcal{M})$.
The D-bracket is mapped to the derived bracket of Poisson brackets,
\begin{eqnarray}
&& j:\vbracket{X}{Y}_D \mapsto - \{ \{ \tilx, H \}, \tily \}.
\label{operation13}
\end{eqnarray}
where $X, Y \in \mathfrak{X}(\wM)$.
A derived bracket construction of a D-bracket has been analyzed in \cite{Deser:2014mxa, Deser:2016qkw, Heller:2016abk}.
Therefore a Vaisman algebroid structure on $T\wM$ is mapped to the mapping space
$\Map(N, T\wM)$ with a symplectic form $\omega$, and operations are calculated by \eqref{operation11} and \eqref{operation13}.

From the above correspondence, a cotangent path is mapped to
\begin{eqnarray}
&& j: (\gamma, \alpha) \mapsto (\varphi(t), \tila(t)).
\end{eqnarray}
where a metric $\eta$ naturally identify $T\wM$ to $T^*\wM$.
Quantities $\phi, \psi, \beta, \zeta$ and tilde quantities in a pre-rackoid which satisfy differential equations \eqref{differentialequations} are 
formally described by exponential maps, which are 
Wilson lines
\begin{eqnarray}
j:(\phi, \beta) &\mapsto& \mathrm{P} \exp \tilx = \mathrm{P} \exp \int_{I_t} X^I(\varphi(\sigma)) A_I(\sigma).
\end{eqnarray}
Similar to the Courant algebroid case, 
the equivalent but more familiar construction for quantization of a Courant algebroid using a topological sigma model is an AKSZ type sigma model formulation.

\section{Conclusion and discussions} \label{sct:conclusion}
In this paper, we studied a global aspect of the doubled geometry in 
DFT through the coquecigrue problem of the Vaisman algebroid.

We first focus on the global structure associated with the Courant algebroid.
A global, group-like structure corresponding to the Leibniz algebroid is a rackoid,
which is a groupoid-like generalization of a rack.
A Leibniz algebroid appears in the tangent bundle of a rackoid as an
infinitesimal counterpart of the global structure.
The cotangent path rackoid proposed in \cite{Laurent-Gengoux:2018zoz} provides the standard Courant algebroid
as a tangent Leibniz algebroid.

With these results at hand, we studied a generalization of the cotangent path rackoid that gives rise to the
Vaisman algebroid as an infinitesimal algebroid.
It is obvious that the Vaisman algebroid fail to satisfy the Leibniz identity in the Courant algebroid.
Since the Leibniz identity is encoded into the self-distributivity of the 
rack product, an integration of the Vaisman algebroid is given by a rackoid type structure without the self-distributivity.
We called this structure the pre-rackoid.
A crucial ingredient for the construction of a pre-rackoid 
is the doubled cotangent path that is defined by the doubled foliations of the para-Hermitian manifold.
With this structure, we defined the pre-rack product by \eqref{eq:pre-rack_product}.
Due to the intermediate shift between different leaves of the doubled foliations, the self-distributivity of the pre-rack product
is explicitly broken.

This picture is consistent with DFT.
One remembers that the strong constraint in DFT picks up a leaf of the foliations as a physical spacetime.
In this case, the pre-rack product is restricted only on a leaf and the
self-distributivity is trivially recovered.

We next focused on a direct approach on the integration of the Courant and
the Vaisman algebroid.
We introduced the formal exponential map of the adjoint action or the bracket
in algebroids. The resulting structures lead to notions of 
formal rackoid and pre-rackoids.
A formal rackoid together with the metric bundle rackoid structure
defines an integration of the Courant algebroid.
We showed that these notions are generalized to the formal metric pre-rackoid.
The D-bracket in the Vaisman algebroid is exponentiated, providing an
example of the pre-rack product.
Compare to the heuristic approach by the explicit examples of the
(pre-)rack product, the approach based on the a formal integration of the algebroid will
help to understand an intuitive feature of the (pre-)rackoid.
The idea is familiar to a formal deformation of an algebra and a perturbative calculation of a quantum theory.

In the end of the discussion, we exhibited another realization of a (pre)-rackoid.
We introduced a three-dimensional topological sigma model, the so-called
Courant sigma model.
This provides a natural arena for a realization of the Courant
algebroid.
We showed that the structure of the formal rackoid is explicitly
implemented by an exponential map of fields in the Courant sigma model.
This is physically interpreted as Wilson lines associated with gauge
fields. 
The construction is generalized to a topological double sigma model of the Vaisman algebroid and the pre-rackoid.
These examples give physical applications of the (pre-)rackoid.

In this paper, we studied a global structure associated with the Vaisman
algebroid from the several viewpoints.
The underlying geometry is the doubled geometry in DFT.

A pre-rackoid structure is related to a global gauge structure of DFT.
Quantizations of topological sigma models will provide quantum version of doubled geometry, which is important for analysis of quantum T-duality.

It is known that DFT contains various solutions that are not the
ones in supergravity \cite{Berkeley:2014nza, Berman:2014jsa,
Bakhmatov:2016kfn, Lust:2017jox, Kimura:2018hph}. 
They are called the non-geometric solutions.
The global aspects of doubled geometry discussed here will be helpful to understand
these non-geometric nature of spacetimes in string theory.
We noted that the strong constraint in DFT is 
represented by the quantum Yang-Baxter equation for the rack action.
It is well-known that the quantum and the classical Yang-Baxter
relations have deep connections with the integrable systems.
Indeed, the classical Yang-Baxter deformation is a key ingredient of the
integrability of string theory \cite{Klimcik:2002zj, Klimcik:2008eq,
Delduc:2013qra}.
We will come back to these issues in future studies.

\subsection*{Acknowledgments}
The authors would like to thank H.~Mori and K.~Shiozawa for useful
discussions.
The work of N.~I. is supported by the Research Promotion Program for
Acquiring Grants in-Aid for Scientific Research (KAKENHI).
The work of S.~S. is supported by Grant-in-Aid for Scientific
Research (C), JSPS KAKENHI Grant Number JP20K03952.
\\
\\
\paragraph{Data Availability}
The data that supports the findings of this study are available within the article.

\end{document}